\documentclass[twocolumn,amsmath,amssymb,floatfix,superscriptaddress,prl]{revtex4-2}

\usepackage{bm,graphicx,url,epsf,color}
\usepackage{amssymb}
\usepackage{amsthm}
\usepackage{xcolor}
\usepackage{comment}
\usepackage{relsize}
\usepackage{xspace}
\usepackage[normalem]{ulem}
\usepackage[super]{nth}
\usepackage{braket}
\usepackage{soul}
\usepackage{textcomp}
\usepackage[T1]{fontenc}

\graphicspath{{graphics/}}

\usepackage[colorlinks=true,linkcolor=blue,citecolor=blue]{hyperref}

\newcommand{\normord}[1]{:\mathrel{#1}:}

\newcommand{\bd}[1]{{\boldsymbol{#1}}}

\let\origaddcontentsline\addcontentsline

\begin{document}

\title{Universal crossover in the three-channel charge Kondo model at high transparency}

\author{Nicolas Paris}
\affiliation{Sorbonne Universit\'e, CNRS, Laboratoire de Physique Th\'eorique de la Mati\`ere Condens\'ee, LPTMC, F-75005 Paris, France}
\affiliation{Universit\'e Paris Cit\'e, CNRS,  Laboratoire  Mat\'eriaux  et  Ph\'enom\`enes  Quantiques, 75013  Paris,  France}

\author{Nicolas Dupuis}
\affiliation{Sorbonne Universit\'e, CNRS, Laboratoire de Physique Th\'eorique de la Mati\`ere Condens\'ee, LPTMC, F-75005 Paris, France}

\author{Christophe Mora}
\affiliation{Universit\'e Paris Cit\'e, CNRS,  Laboratoire  Mat\'eriaux  et  Ph\'enom\`enes  Quantiques, 75013  Paris,  France}

\graphicspath{{./figures/}}

\begin{abstract}
Quantum impurity models provide a central framework for correlated electron physics, with quantum dots enabling controlled experimental realizations. While their weak-coupling behavior is well understood through mappings to Kondo Hamiltonians, the opposite regime of highly transparent contacts has lacked a controlled theoretical description. Using the functional renormalization group (FRG), we resolve this regime for the three-channel charge Kondo device of Ref.~\cite{iftikhar2018}, benchmarking against conformal field theory by reproducing the universal zero-frequency conductance and, crucially, going beyond it to obtain the full frequency crossover of the conductance and the full temperature crossover of the impurity entropy, together with a continuous line of fixed points for interacting leads. These results establish FRG as a powerful nonperturbative tool for quantum impurity problems in regimes inaccessible to conventional approaches, with direct implications for mesoscopic experiments.
\end{abstract}

\maketitle

\paragraph{Introduction.}Quantum impurity physics provides a window into how a small quantum system can dramatically influence a much larger environment. The classic example is the Kondo effect, where a single magnetic impurity in a metal produces a collective many-body state that reshapes electronic transport~\cite{hewson1993kondo,nozieres1974fermi}. Multichannel extensions reveal exotic non-Fermi-liquid behavior and quantum criticality that have become cornerstones of modern condensed matter theory~\cite{affleck1991critical,affleck1992relevance,affleck1993exact,ludwig1994exact}.
Beyond their conceptual importance, impurity models have become central to nanoscale experiments across a variety of platforms, ranging from scanning tunneling spectroscopy on magnetic adatoms~\cite{madhavan1998tunneling,li1998kondo,zhao2005controlling,zhang2013temperature,bork2011tunable,trishin2023tuning,bagchi2024probing} to semiconductor quantum dots~\cite{kouwenhoven2001revival,goldhaber1998kondo,goldhaber1998kondoPRL,cronenwett1998tunable,nygaard2000kondo,sasaki2004enhanced,ji2000phase,van2000kondo,sasaki2000kondo,simmel1999anomalous,liang2002kondo,park2002coulomb,jeong2001kondo,craig2004tunable,potok2007observation,chorley2012tunable,keller2015universal}, which can be tuned with unprecedented control.
These developments have opened the door to realizing exotic correlated states with potential connections to fractionalized excitations and anyonic physics~\cite{affleck1991universal,han2022,lopes2020anyons,komijani2020isolating,lotem2022manipulating,gaines2025spin}. A central challenge is to understand how universality survives when the standard mapping to Kondo Hamiltonians breaks down.

\begin{figure}
    \centering
    \includegraphics[width=8.7cm]{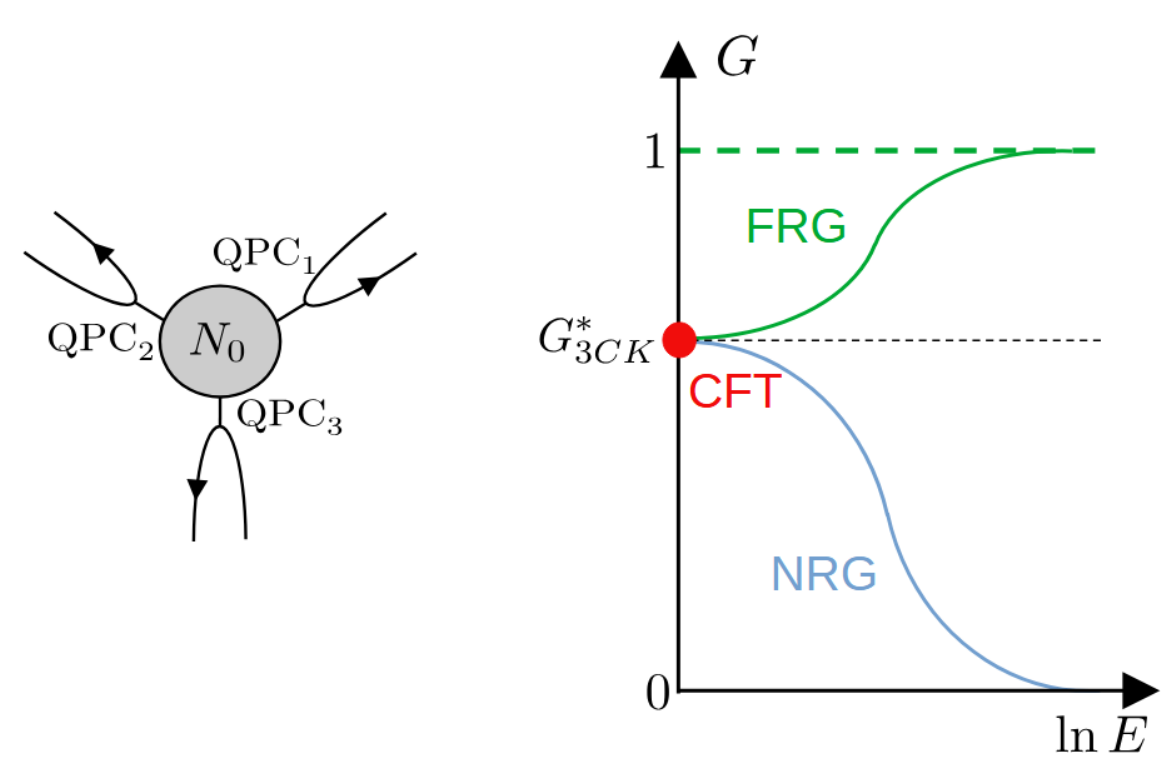}
    \vspace{-7mm}

\caption{\label{fig:exp_device}(Left) Experimental device from Ref.~\cite{iftikhar2018}: a metallic island (the Kondo impurity) with gate-induced offset charge $N_0$ is coupled to three quantum Hall edge channels via quantum point contacts. (Right) Schematic conductance $G$ as a function of energy scale $E$ (temperature or frequency) for low (blue) and high (green) transparencies. Both regimes yield the same universal zero-frequency conductance $G_{\rm 3CK}^*=2\sin^2(\pi/5)$ (red point), as predicted by conformal field theory~\cite{affleck2001,yi1998,yi2002,bao2017quantum}.}
\end{figure}

Metallic islands coupled to quantum Hall edge channels provide an ideal platform for exploring this problem. In the low-transparency regime, tunneling involves only two charge states, and the system maps onto charge versions of multichannel Kondo models~\cite{matveev1991quantum,matveev1995,furusaki1995}. In this regime, numerical renormalization group (NRG) calculations are quantitatively accurate~\cite{bulla2008numerical,mitchell2014generalized} and have been confirmed experimentally in one- and two-island geometries~\cite{han2022,iftikhar2015,iftikhar2018,piquard2023observing,pouse2023quantum,karki2023z}. A hallmark of this regime is the universality of the low-energy behavior, established in the two-channel Kondo model through theory~\cite{sela2011exact,mitchell2012exact,mitchell2016} and experiment~\cite{iftikhar2018}. However, at high contact transparency the mapping fails, as many charge states contribute simultaneously. Despite this breakdown, experiments on the three-channel case suggest that universality persists, with the same zero-frequency conductance observed in both limits~\cite{iftikhar2018}, as illustrated in Fig.~\ref{fig:exp_device}.

In this Letter, we provide a theoretical framework for this strongly coupled regime. Focusing on the charge three-channel Kondo model (3CK), we show that its effective description maps onto a $(0+1)$-dimensional sine-Gordon-like model~\cite{yi1998,yi2002,affleck2001}, which we solve using a nonperturbative functional renormalization group (FRG) approach~\cite{daviet2019,jentsch2022,daviet2023}. We obtain the full universal crossover of the conductance and impurity entropy, controlled by a single energy scale $T^*$. The corresponding universal scaling functions are precisely the quantities probed experimentally, where data are collapsed as functions of frequency or temperature scaled by $T^*$ and compared to theoretical curves. Our results reproduce the exact zero-frequency conductance of conformal field theory (CFT) with 2\% accuracy, establishing the universality of the low-energy physics across all transparencies~\cite{yi1998,yi2002,bao2017quantum}. Extending the analysis to interacting leads, we reveal a continuous line of nonperturbative fixed points, parameterized by the Luttinger parameter $K$ and connected to pseudo-gap Kondo physics~\cite{kim2003pseudo,hopkinson2005quantum}. These findings demonstrate that FRG can access impurity regimes beyond the reach of standard methods, providing quantitative predictions for ongoing experiments.

\paragraph{Three-channel charge Kondo model.} We consider the three-channel charge Kondo geometry depicted in Fig.~\ref{fig:exp_device} and realized experimentally in Refs.~\cite{iftikhar2015,iftikhar2018}. When the single-particle level spacing in the island is the smallest energy scale in the system, the island can be treated as a zero-dimensional object. In this limit, the island and the three leads are described by a one-dimensional model~\cite{matveev1995,furusaki1995} with Hamiltonian
\begin{equation}\label{eq:H_0}
\hat H_0=-i v_F \sum_{j=1}^3\int_{-\infty}^{+\infty} dx \left(\hat \psi^\dagger_{j,R}\partial_x\hat \psi_{j,R}-\hat \psi^\dagger_{j,L}\partial_x\hat \psi_{j,L}\right),
\end{equation}
where $\hat \psi_{j,R/L}$ annihilates a right/left-moving spinless electron in channel $j$, and $v_F$ is the Fermi velocity (we set $\hbar=k_B=1$). The region $x>0$ corresponds to the three pairs of
counter-propagating spin-polarized quantum Hall edge states, while the region $x<0$ represents electronic states inside the metallic island. Using standard bosonization techniques~\cite{vondelft1998,giamarchi2003,dupuis2023field}, Eq.~\eqref{eq:H_0} becomes
\begin{equation}\label{eq:H_0bos}
    \hat H_0=\sum_{j=1}^{3}\int_{-\infty}^{+\infty} dx\,\frac{v}{2\pi}\left[\frac{1}{K}(\partial_x\hat \phi_j)^2+K(\partial_x\hat \theta_j)^2\right],
\end{equation}
where $v$ is the velocity of the density mode.
The special value $K=1$ of the Luttinger parameter (with $v=v_F$) describes non-interacting electrons, and is realized with integer quantum Hall channels in Ref.~\cite{iftikhar2018}. The operator $\hat \phi_j$ is canonically conjugate to the current operator $\partial_x\hat \theta_j/\pi$. The charge density in lead $j$ is given by $-\partial_x\hat \phi_j/\pi$.  The total charge accumulated on the island,
$\hat N=-(1/\pi)\sum_j\hat \phi_j(0)$, enters the charging energy
\begin{equation}\label{eq:H_C}
\hat H_C=E_C(\hat N-N_0)^2 , 
\end{equation}
where $N_0$ is the gate-induced offset charge and $E_C=e^2/2C$ with $C$ the island effective capacitance.
The last contribution to the Hamiltonian accounts for the electron backscattering amplitude $\tilde r_j$ at each quantum point contact (QPC) between the island and the leads,
\begin{equation}
\hat H_{\rm QPC}=v_F\sum_{j=1}^3\tilde r_j (\hat \psi^\dagger_{j,R}(0)\hat \psi_{j,L}(0)+ {\rm H.c.}).
\end{equation}
In bosonized form, this term reads~\cite{refSM}
\begin{equation}\label{eq:H_int}
\hat H_{\rm QPC}=-\sum_{j=1}^3 \frac{D\tilde r_j}{\pi}\cos (2\hat \phi_j(x=0)),
\end{equation}
where $D$ denotes the electron bandwidth.

For low transparencies ($\tilde r_j\gg 1$), the fields $\hat \phi_j(0)$ are pinned to zero modulo $\pi$, corresponding to integer island charge. At $N_0=1/2$, the charging energy [Eq.~\eqref{eq:H_C}] selects the two charge states $N=0,1$, which form a pseudospin-$1/2$. At temperatures $T\ll E_C$, the other charge states can be ignored and the Hamiltonian maps to a Kondo model with three non-interacting conduction channels for $K=1$. As the temperature decreases, the Kondo interaction progressively screens the pseudospin. Initially suppressed due to the low transparency of the contacts, the conductance through the island increases as the temperature is lowered, approaching the zero-temperature 3CK universal value $G^*_{3\mathrm{CK}}= 2 \sin^2 (\pi/5)$~\cite{yi1998}, which reflects partial screening of the impurity, see Fig.~\ref{fig:exp_device}. This result is consistent with experimental observations~\cite{iftikhar2018}.

In the opposite limit of high transparency ($\tilde r_j\ll 1$), many charge states contribute simultaneously. Integrating out the bosonic fields away from the contacts ($x\neq0$) and the gapped charge mode~\cite{refSM} yields the Euclidean action (for $N_0=1/2$),
\begin{align}
    \mathcal S[\bd \varphi]={}&\frac{1}{2\pi K}\sum_{\omega_n}|\omega_n|\,|\bd \varphi(i\omega_n)|^2 \nonumber\\
    &-\sum_{j=1}^{3} r_j\int_0^\beta d\tau \cos\Bigl(\bd G_j\cdot \bd \varphi(\tau)-\frac{\pi}{3}\Bigr) ,
\label{eq:action_QBM}
\end{align}
where $\bd \varphi(\tau)$ is a two-component field linearly related to $\phi_{1,2,3}$. The imaginary time $\tau\in[0,\beta]$, with $\beta=1/T$, and $\omega_n=2n\pi/\beta$ ($n\in\mathbb{Z}$) is a Matsubara frequency.
The reciprocal lattice vectors $\bd G_1=(1,-1/\sqrt{3})$, $\bd G_2=(0,2/\sqrt{3})$, and $\bd G_3=-\bd G_1-\bd G_2$ define a triangular potential landscape. The effective reflection coefficients are
\begin{equation}
r_j=\frac{D}{\pi}\left(\frac{3e^{\gamma/2} K E_C}{\pi D}\right)^{K/3}\tilde r_j ,
\end{equation}
with $\gamma$ the Euler constant~\cite{refSM}. In the following, we focus on the symmetric case where all reflection coefficients are equal, $r_j\equiv r$. The action~\eqref{eq:action_QBM}, supplemented by the ultraviolet (UV) energy cutoff $D$, also describes the quantum Brownian motion of a particle with coordinate $\bd \varphi$ in a two-dimensional periodic potential~\cite{yi1998,yi2002,affleck2001}. 
Although Eq.~\eqref{eq:action_QBM} is not, strictly speaking, a Kondo model, its low-energy fixed point, which is discussed below, coincides with that of the three-channel charge Kondo model~\footnote{In the conventional charge Kondo formulation (low-transparency regime), the renormalization-group flow from the zero-transparency regime toward the 3CK fixed point takes place in the $SU(2)_3$ sector. In contrast, the flow from the fully-transparent regime to the 3CK fixed point occurs in the $SU(2)_3/U(1)$ sector~\cite{affleck2001,bao2017quantum}.}.

\paragraph{FRG approach.} The reflection coefficient $r$ has a scaling dimension $1-2K/3$, hence it is relevant for $K<3/2$. While perturbation theory applies near $K=3/2$, a nonperturbative treatment is required otherwise. We employ the nonperturbative FRG~\cite{dupuis2021}, a modern formulation of Wilson’s RG, which provides the effective action $\Gamma[\boldsymbol{\phi}]$ (or Gibbs free energy in statistical physics terminology) from an exact functional flow equation obtained by progressively integrating out  high-energy degrees of freedom~\cite{refSM}. The effective action 
\begin{equation}
	\Gamma[\bd\phi]	=-\ln Z[\bd {J}]+\int d\tau\, \bd {J}\cdot \bd\phi , 
\end{equation}
is defined as the Legendre transform of the free energy
\begin{equation}
	-\ln Z[\bd {J}]=-\ln\int \mathcal D\bd \varphi \,e^{- S[\bd \varphi]+\int d\tau\, \bd{J}\cdot \bd \varphi} ,
\end{equation}
and $\bd\phi(\tau)=\delta\ln  Z[\bd J]/\delta \bd {J}(\tau)=\langle \bd \varphi(\tau)\rangle$ is the expectation value of the field. 

The first quantity we compute is the effective potential $U(\bd\phi)=\frac{1}{\beta}\Gamma[\bd\phi]|_{\bd\phi\,{\rm const}}$, which is obtained from the effective action evaluated in a time-independent field configuration. The free energy $F=U(\bd\phi=0)$ is given by the effective potential for a vanishing field, and the entropy is simply $S=-\partial U(0)/\partial T$. The second quantity of interest is the two-point vertex, defined as the second functional derivative of the effective action: $\Gamma^{(2)}_{ij}(\tau,\tau')=\delta^2\Gamma[\bd\phi]/\delta\phi_i(\tau) \delta\phi_j(\tau')|_{\bd\phi=0}$. Its Fourier transform $\Gamma^{(2)}_{ij}(i\omega_n)$ is the inverse of the correlation function $\langle \varphi_i(i\omega_n) \varphi_j(-i\omega_n) \rangle$. After analytic continuation to real frequency, it determines the dimensionless conductance matrix $G_{ij}(\omega,T)=G(\omega,T) (\delta_{ij}-\frac{1}{3})$, which gives the current flowing through QPC$_i$ in response to a voltage applied to the electrode connected to QPC$_j$. The conductance $G(\omega,T)$ is given by
\begin{equation}
  G(\omega,T) = {\rm Re} \left\{ \frac{|\omega_n|}{2\pi K}\langle\bd \varphi(i\omega_n)\cdot \bd \varphi(-i\omega_n)\rangle \Bigl|_{\omega_n\to -i\omega+0^+} \right\}  
\end{equation}
in units of $e^2K/h$. The functional flow equation satisfied by the effective action cannot be solved exactly. We rely on the Blaizot--M\'endez-Galain--Wschebor approximation, which allows us to obtain both the entropy and the frequency-dependent conductance~\cite{refSM}.

\begin{figure}
    \centering
    \includegraphics[width=8.7cm]{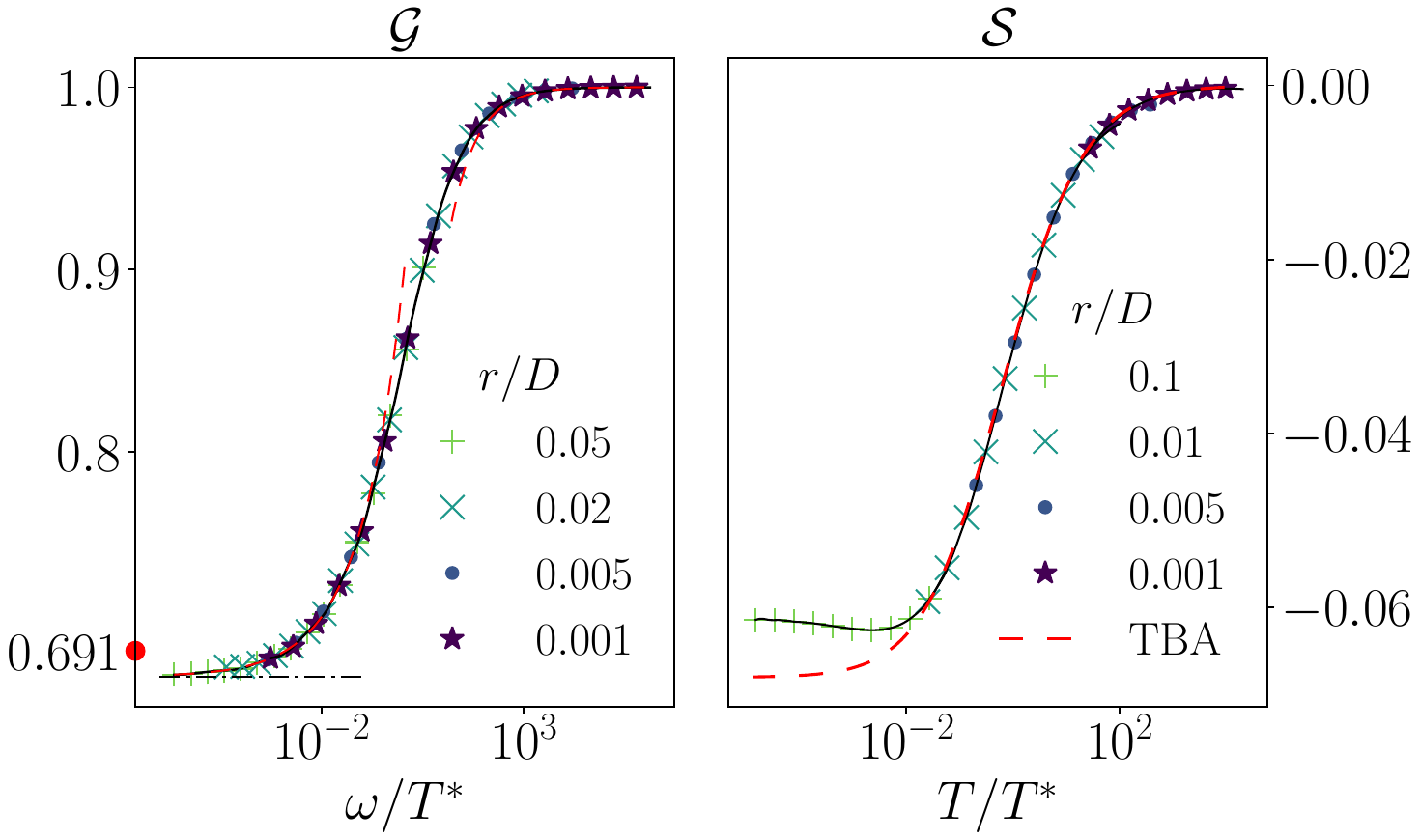}
    \vspace{-7mm}
    \caption{\label{fig:universal_mobility} (Left) Universal conductance (in units of $e^2/h$)
    as a function of rescaled frequency $\omega/T^*$ for several $r$ values at $K=1$. The asymptotic behavior at low-frequency~\eqref{eq:low_freq} and high-frequency~\eqref{eq:high_frequency_asymptotics} is shown by (red) dashed lines, while the horizontal dash-dotted line marks the extrapolated zero-frequency limit. The red point marks the exact CFT result. Curves collapse onto a single line (the black solid line is a guide to the eye) in the wide-band limit $T^*/D\ll 1$.
    (Right) Universal entropy crossover at $K=1$, compared with exact thermodynamic Bethe ansatz (TBA) results. }
\end{figure}

\paragraph{3CK fixed point ($K=1$).}
For noninteracting leads, the zero-temperature conductance approaches unity at high frequency and decreases strongly at low frequency, reflecting the renormalization of QPC transparency (Fig.~\ref{fig:universal_mobility}). While the conductance typically depends on the UV energy cutoff $D$ and the reflection coefficient $r$, in the wide-band limit it exhibits universal scaling as a function of the rescaled frequency $\omega/T^*$,
\begin{equation}
	G(\omega,T=0) = \mathcal G \left( \frac{\omega}{T^*} \right) , 
\label{scalingG}
\end{equation}
where $\mathcal G$ is a universal function (Fig.~\ref{fig:universal_mobility}). Here, $T ^*$ is an emergent low-energy scale that characterizes the crossover between high- and low-frequency behaviors. Its dependence on microscopic parameters can be determined from a perturbative analysis near the ballistic regime $r=0$, 
\begin{equation}\label{eq:T_star}
	\frac{T^*}{D}\propto \left(\frac{r}{D}\right)^3,
\end{equation}
in agreement with FRG~\cite{refSM}. In practice, we define $T^*$ such that $G(\omega=T^*,T=0)$ lies halfway between $1$ and the zero-energy conductance $G^*_{\rm 3CK}=G(\omega=0,T=0)=\mathcal G(0)$. 

The universal function $\mathcal G$ is known in two different limits. In the high-frequency limit $\omega\gg T^*$, it can be obtained perturbatively~\cite{refSM}, 
\begin{equation}\label{eq:high_frequency_asymptotics}
	1-\mathcal G \left(\frac{\omega}{T^*} \right)\underset{\omega\gg T^*}{\propto} \left(\frac{\omega}{T^*}\right)^{-2/3} ,
\end{equation}
which agrees with the FRG calculation (Fig.~\ref{fig:universal_mobility}).
In the low-frequency limit $\omega\ll T^*$, assuming that the low-energy behavior is governed by the 3CK fixed point, regardless of transparency, we can use the results of CFT~\cite{affleck2001,yi1998,yi2002,bao2017quantum}, 
\begin{equation}\label{eq:low_freq}
	\mathcal G \left( \frac{\omega}{T^*} \right) \underset{\omega\to 0}{=}G^*_{\rm 3CK}+c\, \left( \frac{\omega}{T^*}\right)^\nu,
\end{equation}
where $G^*_{\rm 3CK}=2\sin^2(\pi/5)$ and $\nu=2/5$ is the exponent of the leading irrelevant operator. The FRG reproduces both values with remarkable precision (Table~\ref{tab:quantitative_results}). In addition to providing a benchmark for assessing the accuracy of our method, the case $K=1$ also illustrates its ability to access real-frequency dynamical quantities such as $G(\omega)$, which are difficult to compute with other techniques.

The residual impurity entropy exhibits similar scaling in the large bandwidth limit $T^*\ll D$, 
\begin{equation}
	\Delta S(T) = S(T) - S(T)\bigl|_{r=0} = 
	\mathcal S\left(\frac{T}{T^*}\right) , 
\label{scalingS}
\end{equation}
with $\mathcal S$ a universal function. At high temperatures, transport through the QPCs is nearly ballistic, and the residual entropy remains close to zero. It decreases with temperature and takes negative values, larger than $-\ln 2$, indicating that the impurity is partially screened (complete screening would give $\Delta S(0)=-\ln 2\simeq -0.693$). Whereas the conductance is known exactly only at zero temperature and frequency, the full temperature dependence of the entropy can be obtained from the thermodynamic Bethe ansatz (TBA) equations~\cite{affleck2001,refSM}. The universal scaling function obtained from FRG is nearly identical to the exact result over a broad temperature range $T/T^*\gtrsim 10^{-1}$. Eventually, it deviates from the exact TBA solution in the low-temperature limit, although $\mathcal S(0)$ remains within 9\% of the exact value. The exact result, $\mathcal S(0)=\ln[(1+\sqrt{5})/2\sqrt{3}]$, has been interpreted as a signature of emerging Fibonacci anyons~\cite{andrei1984,tsvelick1985exact,han2022,lopes2020anyons}. Whether the FRG captures these (putative) low-energy excitations remains an open question.
 
\begin{table}[b]
\caption{\label{tab:quantitative_results}
Comparison of FRG and exact results for the zero-temperature and zero-frequency conductance, the leading irrelevant exponent $\nu$, and the zero-temperature impurity entropy.}
\begin{ruledtabular}
\begin{tabular}{ccc}
\textrm{}&
\textrm{FRG}&
\textrm{Exact}\\
\colrule
$G^*_{\rm 3CK}$ & $0.677$ & $2\sin^2(\pi/5)\simeq 0.691$~\cite{yi1998} \\
$\nu$ & 0.408  & 2/5~\cite{yi2002}\\
$\Delta S^*(0)$ & -0.062 & $\ln\left(\frac{1+\sqrt{5}}{2\sqrt{3}}\right)\simeq -0.068$~\cite{affleck2001} \\
\end{tabular}
\end{ruledtabular}
\end{table}

\paragraph{Generalized 3CK fixed points ($K\neq 1$).}

\begin{figure}
	\centering
	\includegraphics[width=8.7cm]{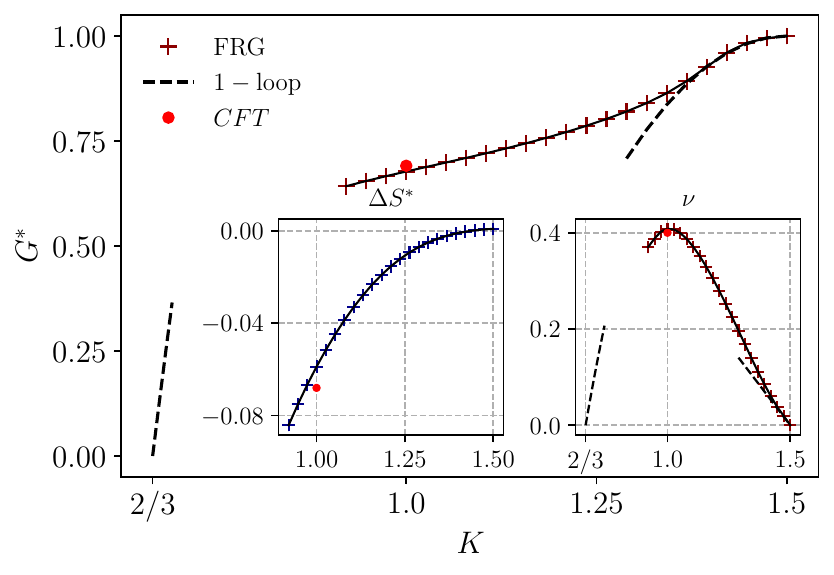}
    \vspace{-8.5mm}
	\caption{\label{fig:general_K} Zero-frequency conductance $G^*(K)$ as a function of $K$, along with the corresponding exponent $\nu(K)$ (right inset). Dashed lines indicate perturbative RG results near $K=3/2$, and near $K=2/3$ obtained from the dual model~\cite{yi1998,yi2002}. The exponent $\nu$ peaks around $K=1$ (right inset). The left inset shows the residual impurity entropy $\Delta S^*(K)$. Solid (black) lines are guides to the eye.}
\end{figure}

Electron interactions in the leads are accounted for by a Luttinger parameter $K\neq1$. In this regime, exact CFT or TBA solutions are unavailable, but perturbative RG analyses~\cite{yi1998,yi2002} predict a rich phase diagram with three distinct regions. For $K>3/2$, the weak-coupling fixed point ($r=0$) is stable, yielding unit conductance at zero temperature and frequency. For $K<2/3$, the strong-coupling fixed point, which can be studied perturbatively in the dual model~\cite{yi1998,yi2002}, is stable and the conductance vanishes. Between these limits, $2/3<K<3/2$, both extremes are unstable, giving rise to a continuous line of intermediate fixed points displaying Kondo-type physics, as in the case $K=1$.

\begin{figure}
	\centering
	\includegraphics[width=8.5cm]{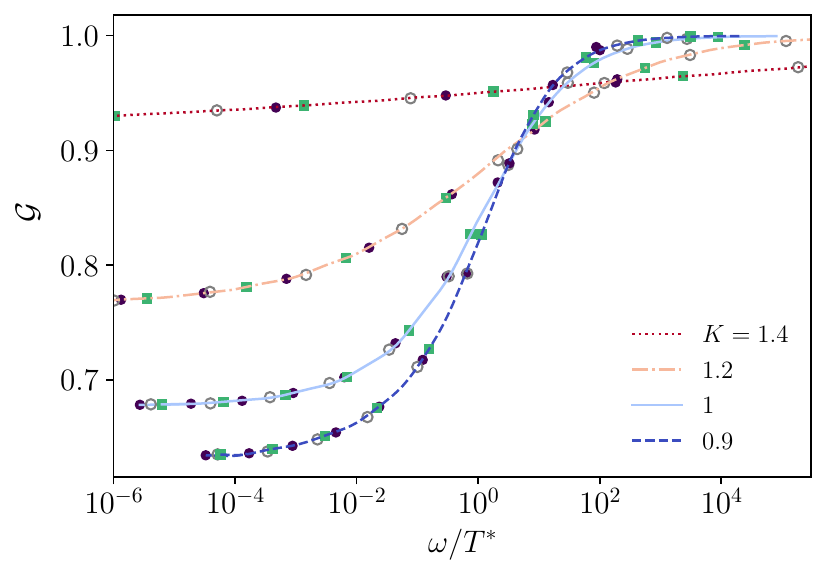}
    \vspace{-5mm}
	\caption{\label{fig:univ_K} Universal conductance scaling function $\mathcal{G}$ for several $K$ values. Markers correspond to $r/D \in \{0.005, 0.01, 0.05\}$. Lines are guides to the eye.}
\end{figure}

Our FRG analysis fully endorses this scenario. The FRG reproduces the ballistic conductance $G(\omega=0,T=0)=1$ for $K>3/2$ and reveals the line of intermediate fixed points that emerges as $K$ decreases. The perturbative RG results near $K = 3/2$ from Ref.~\cite{yi2002} are recovered. The behavior for $2/3 < K < 3/2$ parallels that at $K = 1$: The conductance approaches $1$ at high frequencies and a nonzero value $G^*(K)<1$ in the zero-frequency limit (Fig.~\ref{fig:general_K}). The residual entropy remains close to zero at high temperatures and reaches a finite value $\Delta S^*(K)$ at $T = 0$ (Fig.~\ref{fig:general_K}). The universal scaling forms ~(\ref{scalingG}) and (\ref{scalingS}) of conductance and residual entropy extend beyond the case $K=1$ and remain valid throughout the intermediate region (see Fig.~\ref{fig:univ_K}). Thus, the FRG provides quantitative predictions for the conductance and residual entropy over a wide range of $K$ values, which could be tested in future experiments. Note that the range of $K$ we can study is limited, as integrating the FRG equations becomes increasingly computationally expensive for $K\lesssim 0.9$.

\paragraph{Conclusion.}In summary, we have solved the three-channel charge Kondo model in the quasi-ballistic regime using the functional renormalization group. Our results recover the universal zero-frequency conductance of the three-channel Kondo fixed point with excellent quantitative accuracy and provide the full universal crossover of the conductance and impurity entropy as functions of temperature or frequency. Extending the analysis to interacting channels, we uncovered a continuous line of fixed points that connects smoothly to pseudo-gap Kondo physics in the low-transparency limit. While tuning electron–electron interactions in quantum Hall systems is challenging, recent experiments with multiple islands demonstrate viable routes to enhance them, opening possibilities to test our predictions for $K \ne 1$~\cite{parafilo2024,anthore2018}. More broadly, this work establishes FRG as a powerful nonperturbative approach to quantum impurity problems in regimes inaccessible to conventional methods, with direct implications for future experiments in mesoscopic devices.

\paragraph{Acknowledgements.} We acknowledge fruitful discussions with A. Anthore, F. Pierre, C. Piquard and F. Rose. C.~M. acknowledges funding
from the Agence Nationale de la Recherche under the France 2030 programme, reference
ANR-22-PETQ-00122 (EQUBITFLY).

\renewcommand{\addcontentsline}[3]{}

\let\addcontentsline\origaddcontentsline  

\clearpage
\newpage 

\onecolumngrid

\begingroup  

\setcounter{equation}{0}
\setcounter{figure}{0}
\setcounter{table}{0}
\setcounter{page}{1}
\renewcommand{\theequation}{S\arabic{equation}}
\renewcommand{\theHequation}{S\arabic{equation}}
\renewcommand{\thefigure}{S\arabic{figure}}	
\renewcommand{\thetable}{S\arabic{table}}	

\renewcommand{\bibnumfmt}[1]{[S#1]}
\renewcommand{\citenumfont}[1]{S#1}

\centerline{\bf Supplemental Material: Three-channel charge Kondo model at high transparency}
\medskip
\centerline{Nicolas Paris, Nicolas Dupuis, and Christophe Mora}

\bigskip

In this Supplemental Material, we derive the effective action used to calculate the conductance and the entropy of the experimental setup discussed in the main text. We briefly review the functional renormalization group (FRG) approach, including the local potential approximation (LPA)  and the Blaizot--M\'endez-Galain--Wschebor (BMW) approximation scheme, and we discuss the numerical solution of the flow equations. 

\tableofcontents

\section[I. Model and bosonization]{I. Model and bosonization}

The experimental setup under consideration~\cite{sm_iftikhar2015,sm_iftikhar2018} consists of a metallic island coupled to three one-dimensional quantum Hall edge channels. When the level spacing of the states on the island is the smallest energy scale in the system, the finite volume of the island can be neglected. Under these conditions, the island and the three channels can be described by a one-dimensional model~\cite{sm_matveev1995,sm_furusaki1995} with the Hamiltonian $\hat H_0+\hat H_C+\hat H_{\rm QPC}$, 
\begin{equation}
	\begin{split} 
		&\hat H_0=-i v_F \sum_{j=1}^3\int_{-\infty}^{+\infty} dx\, \bigl(\hat \psi^\dagger_{j,R}\partial_x \hat \psi_{j,R}-\hat \psi^\dagger_{j,L}\partial_x\hat \psi_{j,L}\bigr) , \\
		&\hat H_C=E_C(\hat N-N_0)^2 , \\
		&\hat H_{\rm QPC}=v_F\sum_{j=1}^3\tilde r_j (\hat \psi^\dagger_{j,R}(0)\hat \psi_{j,L}(0)+ {\rm H.c.}) ,
	\end{split}
\end{equation}
where $\hat \psi_{j,R/L}$ is the annihilation operator for right/left moving (spinless) electrons in the channel $j$ and $v_F$ is the Fermi velocity (we set $\hbar=k_B=1$ throughout). The region $x>0$ corresponds to three pairs of
counter-propagating spin-polarized quantum Hall edge states, 
while the region $x<0$ represents the electronic states of the metallic island. The charging energy $E_C$ of the metallic island is equal to $e^2/2C$, where $C$ is the effective capacitance. $N_0$ is an offset charge controlled by an external gate voltage. The total number of electrons (charge) on the metallic island is given by
\begin{equation}\label{eqapp:N}
	\hat N=\sum_{r=R,L} \sum_{j=1}^3 \int_{-\infty}^0dx\,\hat \rho_{j,r}(x) ,
\end{equation}
where $\hat \rho_{j,r}(x)=\normord{\hat \psi^\dagger_{j,r}(x)\hat \psi_{j,r}(x)}$ is the (normal ordered) density operator. The microscopic reflection amplitude at the quantum point contact (QPC) between the island and the $j$th channel is denoted by $\tilde r_j$.

A convenient formulation of the Hamiltonian can be obtained by using standard bosonization techniques~\cite{sm_vondelft1998,sm_giamarchi2003,sm_dupuis2023field}. The electron operators are expressed in terms of bosonic operators and fermionic ladder operators (Klein factors),
\begin{equation}\label{eq:bosonization}
	\hat \psi_{j,r}(x)=\frac{\hat U_{j,r}}{\sqrt{2\pi \alpha}}e^{irk_Fx-i\left(r\hat \phi_j(x)-\hat \theta_j(x)\right)},	
\end{equation}
where $r=+/-$ for right/left-moving particles and $\alpha=v_F/D$ is a short-distance cutoff, with $D$ the energy bandwidth in the one-dimensional channels. The bosonic operators $\hat \phi_j$ and $\hat \theta_j$ obey the commutation relations $[\hat \phi_i(x),\partial_y \hat \theta_j(y)]=i\pi\delta_{i,j}\delta(x-y)$.
The anticommutation relations satisfied by the Klein factors, $\{\hat U_{j,r},\hat U_{j',r'}\}=2\delta_{jj'}\delta_{rr'}$, ensure that the fermion operators $\hat \psi_{j,r}$ anticommute.

The free Hamiltonian $\hat H_0$ reduces to the Luttinger-liquid Hamiltonian when expressed in terms of bosonic operators. It is straightforward to extend our approach to the case where the electrons in the one-dimensional channels interact. In that case, $\hat H_0$ takes the usual form
\begin{equation}\label{eqapp:H_0bos}
	\hat H_0=\sum_j\int_{-\infty}^\infty  dx\,\frac{v}{2\pi}\left[\frac{1}{K}(\partial_x\hat \phi_j)^2+K(\partial_x\hat \theta_j)^2\right] ,
\end{equation}
where $K$ is the Luttinger parameter and $v$ is the density mode velocity. In the absence of interactions, $K=1$ and $v=v_F$. Using the relation $\hat \rho_{j}(x)=- \partial_x\hat \phi_j(x)/\pi$, we can write the total charge on the island as  
\begin{align}
	\hat N=-\frac{1}{\pi}\sum_j\hat \phi_j(0) ,
\end{align}
where we have assumed $\hat \phi_j(-\infty)=0$. This leads to 
\begin{equation}\label{eqapp:N_bos}
	\hat H_C=E_C\biggl(\frac{1}{\pi}\sum_j\hat \phi_j(0)+N_0\biggr)^2 . 
\end{equation}
The action associated with the electron backscattering at the QPCs takes the form
\begin{equation}
	\hat H_{\rm QPC}=\frac{D}{2\pi} \sum_j \tilde r_j\hat U_{j,+}^\dagger\hat U_{j,-} e^{2i\hat \phi_j(0)} + {\rm H.c.} 
\end{equation}
Since the operators $\hat U^\dagger_{j,r}\hat U_{j,-r}$ and the Hamiltonian commute pairwise, they can be diagonalized simultaneously. Furthermore, unitarity implies that the eigenvalues of $\hat U^\dagger_{j,+}\hat U_{j,-}$ and $\hat U^\dagger_{j,-}\hat U_{j,+}$ are of the form $e^{i\gamma}$ and $e^{-i\gamma}$, respectively. To preserve the symmetry between the three channels when $r_j=r$, we choose $\gamma$ to be independent of $j$. This leads to 
\begin{equation}\label{eqapp:H_intbos}
	\hat H_{\rm QPC}=\frac{D}{\pi} \sum_j\tilde r_j\cos (2\hat \phi_j(0)+\gamma).
\end{equation}

A convenient description of the island can be obtained by integrating out the degrees of freedom away from the QPCs, i.e. those associated with $\hat \phi_j(x\neq0)$. This is most easily done in the functional integral formalism~\cite{sm_NDbook1}. This leads to an effective (0+1)-dimensional Euclidean action $S_0+S_{\rm int}$,
\begin{equation}\label{eqapp:action}
	\begin{split}
		&S_0=\sum_{\omega_n}\sum_j \frac{|\omega_n|}{\pi K}|\phi_j(i\omega_n)|^2 , \\
		&S_{\rm int}=\frac{D}{\pi} \sum_j \tilde r_j  \int _0^{\beta}d\tau \cos(2\phi_j(\tau)+\gamma)+ E_C\int _0^\beta d\tau\biggl(\frac{1}{\pi}\sum_j\phi_j(\tau)+N_0\biggr)^2  ,
	\end{split}
\end{equation}
where $\phi_j(\tau)$ is a bosonic (real) field, $\tau$ is the imaginary time, and $\beta=1/T$ is the inverse temperature. We now denote by $\phi_j(\tau)\equiv \phi_j(x=0,\tau)$ the field at $x=0$ and by $\phi_j(i\omega_n)$ its Fourier transform, with $\omega_n$ a bosonic Matsubara frequency. The energy bandwidth $D$ emerges as the natural frequency cutoff. The phase $\gamma$ can be eliminated by shifting the field, $\phi_j(\tau)\to \phi_j(\tau)-\gamma/2$, which leads to a simple redefinition of $N_0$. In the following, we set $\gamma=\pi$. With this choice, the three-channel Kondo (3CK) fixed point is reached for half-integer values of $N_0$. 

Next, we perform a change of variables
\begin{equation}\label{eq:change_var}
	\begin{pmatrix}
		\phi_1\\
		\phi_2\\
		\phi_3
	\end{pmatrix}
	=-\begin{pmatrix}
		\frac{1}{\sqrt{6}}&\frac{1}{2}&\frac{1}{\sqrt{12}}\\
		\frac{1}{\sqrt{6}}&0&-\frac{1}{\sqrt{3}}\\
		\frac{1}{\sqrt{6}}&-\frac{1}{2}&\frac{1}{\sqrt{12}}\\
	\end{pmatrix}
	\begin{pmatrix}
		\varphi_C\\
		\varphi_1\\
		\varphi_2
	\end{pmatrix} ,
\end{equation}
defined by an orthogonal transformation (up to a factor $1/\sqrt{2}$). 
The field $\varphi_C=\pi \sqrt{2/3}N$ describes fluctuations of the total charge $N$ on the island. The fields $\varphi_1$ and $\varphi_2$ are defined up to an arbitrary rotation in the plane perpendicular to $\varphi_C$. This degree of freedom corresponds to a rotation of the vectors ${\bf G}_i$ defined in the following and is irrelevant for physical quantities. With the additional shift $\varphi_C\to  \varphi_C+\pi \sqrt{2/3}N_0$, the action reads
\begin{align}
	S=&\frac{1}{2\pi K}\sum_{\omega_n}\Bigl[|\omega_n||\varphi_1(i\omega_n)|^2+|\omega_n||\varphi_2(i\omega_n)|^2 +|\varphi_C(i\omega_n)|^2\Bigl(|\omega_n|+\frac{3}{\pi} K E_C \Bigr)\Bigr]\nonumber\\
	&-\frac{D}{\pi}\int_0^\beta d\tau\biggl[\tilde r_1\cos\biggl(\varphi_1(\tau)+\frac{1}{\sqrt{3}}\varphi_2(\tau)+\sqrt{\frac{2}{3}}\varphi_C(\tau)+\frac{2\pi}{3}N_0\biggr)+\tilde r_2\cos\biggl(\frac{2}{\sqrt{3}}\varphi_2(\tau)-\sqrt{\frac{2}{3}}\varphi_C(\tau)-\frac{2\pi}{3}N_0\biggr)\nonumber\\
	&+\tilde r_3\cos\biggl(\varphi_1(\tau)-\frac{1}{\sqrt{3}}\varphi_2(\tau)-\sqrt{\frac{2}{3}}\varphi_C(\tau)-\frac{2\pi}{3}N_0\biggr)\biggr] . 
\end{align}
The charge mode $\varphi_C$ is gapped and is little affected by the periodic potential in the weak-backscattering limit $\tilde r_j\ll E_C/D$. Its fluctuations determine the charge fluctuations on the island. To leading order in $\tilde r_j$, and in the limit $T\ll E_C\ll D$,  
\begin{equation}
	\langle (\hat N-N_0)^2 \rangle =
	\frac{3}{2\pi^2} \langle \varphi_C^2 \rangle \simeq 
	\frac{3K}{2\pi^2} \ln\left(\frac{\pi D}{3e^{\gamma/2}KE_C }\right) ,
\end{equation}
with $\gamma$ being the Euler constant. We have used 
\begin{align}\label{eq:phi_c_fluc}
	\langle\varphi_C^2\rangle = \pi KT\sum_{\omega_n}\frac{e^{-\omega^2_n/D^2}}{|\omega_n|+\frac{3}{\pi}KE_C} \simeq K \ln\left(\frac{\pi D}{3e^{\gamma/2}KE_C }\right) \quad (T\ll E_C \ll D) , 
\end{align}
where the exponential factor implements the ultraviolet (UV) energy cutoff. Since $\langle (\hat N-N_0)^2 \rangle\gg 1$ when $E_C \ll D$, many charge states must be taken into account in the large-bandwidth limit. 

In the spirit of a cumulant expansion, the charge mode can be integrated out~\cite{sm_matveev1995,sm_furusaki1995}. Using~(\ref{eq:phi_c_fluc}), we obtain 
\begin{align}	
	\langle e^{i\sqrt{2/3}\varphi_C}\rangle=e^{-\frac{1}{3}\langle \varphi_C^2\rangle}=\left(\frac{3e^{\gamma/2} K E_C}{\pi D}\right)^{K/3}
\end{align}
in the limit $\tilde r_j\to 0$. To leading order in $\tilde r_j$, this yields the following effective action for the field $\bd\varphi=(\varphi_1,\varphi_2)$, 
\begin{equation}\label{eq:action_QBM}
	S=\sum_{\omega_n }  \frac{|\omega_n|}{2\pi K}  |\bd\varphi(i\omega_n)|^2 -\sum_{j=1}^{3}  r_j\int_0^\beta d\tau \cos\Bigl(\bd G_j\cdot \bd \varphi(\tau)-\frac{2\pi}{3}N_0 \Bigr) ,
\end{equation}
where
\begin{equation}
	\bd G_1=\begin{pmatrix}
		-1\\
		-\frac{1}{\sqrt{3}}
	\end{pmatrix}, \hspace{5mm}
	\bd G_2=\begin{pmatrix}
		0\\
		\frac{2}{\sqrt{3}}
	\end{pmatrix}, \hspace{5mm}
	\bd G_3=\begin{pmatrix}
		1\\
		-\frac{1}{\sqrt{3}}
	\end{pmatrix} = - \bd G_1 - \bd G_2 
\end{equation}
and 
\begin{equation} \label{eq:rjdef} 
	r_j=\frac{D}{\pi}\left(\frac{3e^{\gamma/2} K E_C}{\pi D}\right)^{K/3}\tilde r_j . 
\end{equation}
The periodic potential in~(\ref{eq:action_QBM}) possesses the symmetry of a triangular lattice defined by the vectors $\bd a_1=2\pi(1,0)$ and $\bd a_2=2\pi(1/2,\sqrt{3}/2)$ or, equivalently, by the reciprocal vectors $\bd G_1$ and $\bd G_2$. The potential has two minima per unit cell when $N_0$ is half-integer and a single minimum otherwise.
The action~\eqref{eq:action_QBM} also describes the quantum Brownian motion (QBM) of a particle with coordinates $(\varphi_1,\varphi_2)$ moving in a two-dimensional periodic potential~\cite{sm_yi1998,sm_yi2002,sm_affleck2001}. Hereafter, we refer to it as the QBM action.

\section[II. Conductance of the quantum point contacts]{II. Conductance of the quantum point contacts}

The conductance of the experimental setup is defined as a tensor,
\begin{equation}
	I_j=\sum_k G_{jk}V_k , 
\end{equation}
where $I_j$ is the total current flowing through QPC$_j$ to the island, and $V_k$ is the voltage applied to the electrode connected to QPC$_k$. To align with the experimental results of Ref.~\cite{sm_iftikhar2018} (see, in this reference, the Supplemental Material on ``Renormalized channel conductance''), we introduce the individual QPC conductances $G_i=1/R_i$ defined by analogy with the classical circuit shown in Fig.~\ref{fig:circuit}. These conductances can be expressed in terms of the conductance tensor via the Kirchhoff laws,
\begin{equation}
	\sum_j I_j=0 , \quad V_1- \frac{I_1}{G_1}=V_2-\frac{I_2}{G_2}=V_3-\frac{I_3}{G_3}.
\end{equation}
A straightforward calculation yields 
\begin{equation}		
	G_{ij}=G_i\left( \delta_{ij}-\frac{G_j}{\sum_kG_k}\right) .
\end{equation}
If there is no asymmetry between the three channels, all the QPC conductances are equal to $G$ and $G_{ij}=G(\delta_{ij}-1/3)$.

\begin{figure}[b]
	\centering
	\includegraphics[width=6cm]{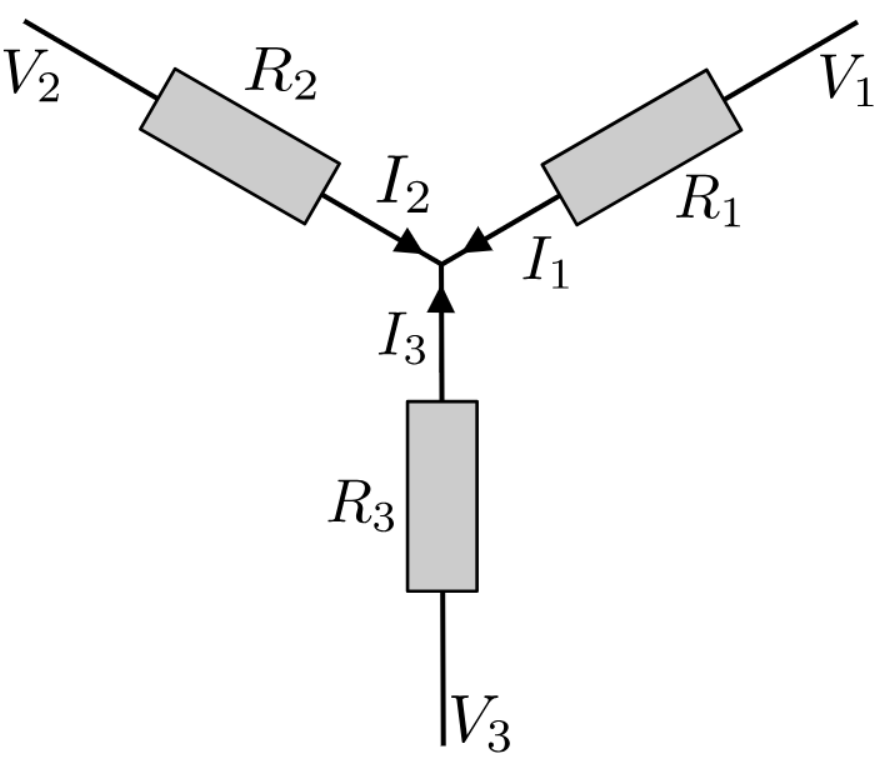}
	\vspace{-0.2cm}
	\caption{\label{fig:circuit}Equivalent classical circuit describing the experimental setup of Ref.~\cite{sm_iftikhar2018}. Each QPC is defined by its own resistance.}
\end{figure}

The conductance tensor can be computed within linear response theory. An applied voltage $V_i(t)$ corresponds to a time-dependent perturbation $\hat H'(t) = \sum_i e V_i(t) \hat N_i$ where $\hat N_i=-\hat \phi_i/\pi$ and $e$ is the electron charge. The linear response is given by 
\begin{equation}
	\langle \hat N_i(t) \rangle = \sum_j \int_{-\infty}^\infty dt' \,  \chi^R_{N_iN_j}(t-t') e V_j(t') = \sum_j \int_{-\infty}^\infty dt' \, \frac{e}{\pi^2}  \chi^R_{\phi_i\phi_j}(t-t') V_j(t'), 
\end{equation} 
where $\chi^R_{AB}(t-t') = i \theta(t-t') \langle [\hat A(t), \hat B(t')] \rangle$
is a retarded response function, and $\hat A(t)$ and $\hat B(t)$ are operators in the Heisenberg picture. Since the expectation value of the current is given by $\langle \hat I_i(t)\rangle =e \frac{d}{dt} \langle \hat N_i(t) \rangle$, we obtain
\begin{equation}
	\langle \hat I_i(t) \rangle = \sum_j \int_{-\infty}^\infty dt' \, \frac{e^2}{\pi^2} \frac{d}{dt}  \chi^R_{\phi_i\phi_j}(t-t') V_j(t'), 
\end{equation} 
which leads to the frequency-dependent conductance 
\begin{equation}
	G_{ij}(\omega) = - \frac{e^2}{\pi^2} {\rm Re} \bigl[ i \omega \chi^R_{\phi_i\phi_j}(\omega) \bigr] . 
\end{equation}
$G_{ij}(\omega)$ can be obtained by analytic continuation of the imaginary-frequency correlation function of the field $\phi_i$,  
\begin{equation}\label{eq:cond_tensor1}
	G_{i j}(\omega)=\frac{e^2}{\pi^2} {\rm Re } \Bigl\{|\omega_n|\langle \phi_i(i\omega_n)\phi_j(-i\omega_n)\rangle \bigr|_{i\omega_n\rightarrow \omega+i0^+}\Bigr\} . 
\end{equation}
After the change of variables~\eqref{eq:change_var}, Eq.~(\ref{eq:cond_tensor1}) can be rewritten in terms of the correlation functions of $\varphi_{1}$ and $\varphi_{2}$ and $\varphi_{C}$. Neglecting all correlation functions involving the gapped field $\varphi_C$, we obtain 
\begin{equation}\label{eq:cond_tensor}
	G_{ij}(\omega)= \frac{e^2}{4\pi^2} {\rm Re } \Bigl\{|\omega_n|\langle \bd G_i\cdot \bd\varphi(i\omega_n) \, \bd G_j\cdot \bd\varphi(-i\omega_n) \rangle\bigr|_{i\omega_n\to \omega+i0^+}\Bigr\}.
\end{equation}
In the symmetric case, where $r_j=r$ and $\langle \varphi_i(i\omega_n) \varphi_j(-i\omega_n)\rangle = (\delta_{ij}/2)\langle \bd\varphi(i\omega_n) \cdot  \bd\varphi(-i\omega_n)\rangle$, the QPC conductances are given by 
\begin{equation}\label{eq:cond_symm}
	G(\omega)= \frac{e^2}{4\pi^2} {\rm Re } \left\{|\omega_n|\langle\bd \varphi(i\omega_n)\cdot \bd \varphi(-i\omega_n)\rangle\bigr|_{i\omega_n\to \omega+i0^+} \right\} .
\end{equation}

\section[III. FRG approach]{III. FRG approach}

In this section, we briefly describe the FRG approach used to obtain the physical properties of the metallic island. Following the standard FRG strategy~\cite{sm_berges2002,sm_delamotte2012,sm_dupuis2021}, we add an infrared regulator term to the QBM action~\eqref{eq:action_QBM},
\begin{equation}
	\Delta S_k[\varphi]=\frac{1}{2} \sum_{i,j=1}^2 \sum_{\omega_n}R_{k,ij}(i\omega_n)\varphi_i(-i\omega_n)\varphi_j(i\omega_n) ,
\end{equation}
where $k$ is a (running) energy scale. The regulator function is defined by 
\begin{equation}
	R_{k,ij}(i\omega_n)= \gamma_1 \delta_{ij}  \frac{|\omega_n|}{\pi K}r\left(\frac{|\omega_n|}{k}\right)+\gamma_2 Z_{k,ij}\omega_n^2r\left(\frac{\omega_n^2}{k^2}\right) .
\end{equation}
The $k$-dependent coefficients $Z_{k,ij}$ are set to zero in the LPA and are defined by~\eqref{eq:Z_Y} in the BMW approximation. We take $r(x)=1/(e^x-1)$ for both the LPA and the BMW approximation. The parameters $\gamma_1$ and $\gamma_2$ are of the order of unity and can be varied to test the sensitivity of our method with respect to the choice of regulator function. 

The partition function associated with the QBM action $S+\Delta S_k=S_0 + S_{\rm int}+\Delta S_k$, 
\begin{equation}
	{\cal Z}_k[\bd J]=\int \mathcal D\bd \varphi \,e^{- S[\bd \varphi]-\Delta S_k[\bd\varphi]+\int_0^\beta d\tau \bd {J}\cdot \bd \varphi}
\end{equation}
is $k$-dependent. The expectation value of the field, 
\begin{equation}
	\bd\phi(\tau)=\frac{\delta \ln {\cal Z}_k[\bd J]}{\delta \bd J(\tau)}=\langle\bd\varphi(\tau)\rangle , 
\end{equation}
can be obtained from a functional derivative with respect to the external source $\bd J$. The scale-dependent effective action,
\begin{equation}
	\Gamma_k[\bd\phi]=-\ln {\cal Z}_k[\bd J]+\int_0^\beta d\tau\, \bd J\cdot \bd \phi -\Delta S_k[\bd\phi] ,
\end{equation}
is defined as a slightly modified Legendre transform that includes the subtraction of $\Delta S_k[\bd\phi]$. Fluctuations are nearly frozen by the regulator term $\Delta S_{k_{\rm in}}$ if the initial value $k_{\rm in}$ of the RG scale $k$ is much larger than the bandwidth $D$ (the natural UV cutoff of the theory). In that case, $\Gamma_{k_{\rm in}}[\bd \phi]\simeq S[\bd \phi]$. On the other hand, the effective action of the original model is given by $\Gamma_{k=0}$ since $R_{k=0}=0$. The nonperturbative FRG approach aims to determine $\Gamma_{k=0}$ from $\Gamma_{k_{\rm in}}$ using Wetterich’s equation~\cite{sm_wetterich1993,sm_ellwanger1994,sm_morris1994},
\begin{equation}\label{eq:wett}
	\partial_t \Gamma_k[\bd\phi]=\frac{1}{2}\mbox{Tr}\left[\partial_t R_k(\Gamma_k^{(2)}[\bd \phi]+R_k)^{-1}\right] , 
\end{equation}
where $\Gamma_k^{(2)}$ is the second functional derivative of $\Gamma_k$ and $t=\ln k$ is a (negative) RG time (which varies from $\ln k_{\rm in}$ to $-\infty$ as $k$ decreases from $k_{\rm in}$ to zero). The trace in Eq.~(\ref{eq:wett}) involves a sum over frequencies as well as field indices.

\subsection[A. LPA]{A. LPA}

In the LPA, the scale-dependent effective action is approximated by
\begin{equation} \label{eq:GamLPA}
	\Gamma_k[\phi]= \sum_{\omega_n}\frac{|\omega_n|}{2\pi K} |\bd\phi(i\omega_n)|^2 + \int_0^\beta d\tau \, U_k(\bd\phi(\tau)) ,
\end{equation}
with the initial condition
\begin{equation}
	U_{k_{\rm in}}(\bd \phi)=-\sum_{j=1}^{3}  r_j \cos \biggl(\bd G_j\cdot \bd \phi - \frac{2\pi}{3}N_0 \biggr) .
\end{equation}
The flow equation preserves the triangular symmetry of the effective potential $U_k(\bd\phi)$, which can be expanded into a Fourier series,
\begin{equation} \label{eq:Uharmonics} 
	U_k(\bd\phi) = \sum_{m,n\in \mathbb{Z}} r_k^{(m,n)}e^{i (m\bd G_1+n \bd G_2)\cdot \bd \phi } ,
\end{equation}
with complex coefficients that satisfy $r_k^{(-m,-n)}=r_k^{(m,n)}{}^*$. The amplitudes of the first-order harmonics, $|r_k^{(1,0)}|$, $|r_k^{(0,1)}|$ and $|r_k^{(1,1)}|$, can be regarded as renormalized reflection coefficients. The flow equation of $U_k(\bd\phi)$ is obtained by inserting the ansatz~(\ref{eq:GamLPA}) into Wetterich's equation,
\begin{equation}\label{eq:LPA}
	\partial_t U_k(\bd \phi)=\frac{1}{2\beta}\sum_{\omega_n}e^{-\omega_n^2/D^2}\,{\rm Tr}[\partial_t R_{k}(i\omega_n)G_k(i\omega_n,\bd \phi)] , 
\end{equation}
The trace is taken with respect to the indices of the $2\times 2$ matrices $R_k(i\omega_n)$ and $G_k(i\omega_n,\bd\phi)$. The propagator $G_k(i\omega_n,\bd\phi)$ is obtained by inverting $\Gamma^{(2)}_k(i\omega_n,\bd\phi)+R_k(i\omega_n)$, where $\Gamma^{(2)}_k(i\omega_n,\bd\phi)$ is the two-point vertex computed in a constant field $\bd\phi$. The factor $e^{-\omega_n^2/D^2}$ in~(\ref{eq:LPA}) implements the UV energy cutoff. In practice, the flow equation~\eqref{eq:LPA} is solved using the dimensionless quantities $\tilde\omega_n=\omega_n/k$ and $\tilde r_k^{(n,m)}=r_k^{(n,m)}/k$.

\subsection[B. BMW approximation]{B. BMW approximation}

The LPA allows us to compute the free energy $F=U_{k=0}(\bd\phi=0)$ and the entropy $S=-\partial F /\partial T$. However, it does not allow for the calculation of the conductance, $G(\omega)$, which requires knowledge of the full frequency dependence of the propagator $G_{k=0,ij}(i\omega_n)=\langle \varphi_i(i\omega_n)\varphi_j(-i\omega_n) \rangle$; see Eq.~(\ref{eq:cond_symm}). A derivative expansion of the effective action beyond the LPA is subject to the same limitations. The propagator $G_{k=0}$, or its inverse the two-point vertex $\Gamma^{(2)}_{k=0}$, can be computed within the BMW approximation~\cite{sm_blaizot2006,sm_blaizot2006PRE,sm_benitez2009,sm_benitez2012}. The latter allows one to close the infinite hierarchy of equations satisfied by the vertices $\Gamma^{(n)}_k$. In its simplest nontrivial implementation, the BMW approximation yields closed equations for the effective potential $U_k(\bd\phi)$ and the two-point vertex $\Gamma^{(2)}_k(i\omega_n,\bd\phi)$ evaluated in a constant field. The latter can be written as 
\begin{equation} 
	\Gamma^{(2)}_{k,ij}(i\omega_n,\bd\phi) = \delta_{ij} \frac{|\omega_n|}{\pi K}+\Delta_{k,ij}(i\omega_n,\bd\phi)+U''_{k,ij}(\bd\phi) , \label{eq:Gamma_BMW}
\end{equation} 
where $\Delta_k(i\omega_n) = \Gamma_k^{(2)}(i\omega_n,\bd\phi) - \Gamma_k^{(2)}(0,\bd\phi)$
is a ``self-energy'', and we use the notation $U''_{k,ij}=\partial_{\phi_i}\partial_{\phi_j}U_k$. 
We anticipate the fact that the Luttinger parameter, i.e. the coefficient of the $|\omega_n|$ term, is not renormalized. At low frequencies, $|\omega_n|\ll k$, the self-energy is expected to be quadratic in frequency. For the sake of numerical accuracy, we write it in the form 
\begin{equation}
	\Delta_{k,ij}(i\omega_n,\bd \phi)=Z_{k,ij}\omega_n^2[1+Y_{k,ij}(i\omega_n,\bd\phi)]\label{eq:Z_Y}
\end{equation}
with the additional constraint
\begin{equation}
	\int d^2 \phi\, Y_{k,ij}(i\omega_n=0,\bd\phi)=0,
	\label{eq:Y_constraint}
\end{equation}
which fixes the value of $Z_{k,ij}$.  

The derivation of the BMW flow equation is standard~\cite{sm_blaizot2006,sm_benitez2009,sm_benitez2012}. Equation~\eqref{eq:LPA} still holds, while $Y_{k,ij}$ satisfies the equation 	
\begin{align}
	\partial_t Y_{k,ij} ={}& \eta_{k,ij}(1+Y_{k,ij}) + \sum_{i_1,i_2} I_{1,k}^{i_1i_2}\partial_{\phi_{i_1}}\partial_{\phi_{i_2}}Y_{k,ij} + \sum_{i_1,i_2,i_3,i_4} \biggl\{ I_{2,k}^{i_1i_2i_3i_4} \biggl[ \frac{Z_{k,ii_3}Z_{k,ji_4}}{Z_{k,ij}}\omega_n^2\partial_{\phi_{i_2}}Y_{k,ii_3}\partial_{\phi_{i_1}}Y_{k,ji_4}\nonumber\\&
	+\frac{Z_{k,ii_3}}{Z_{k,ij}}\partial_{\phi_{i_2}}Y_{k,ii_3}U^{(3)}_{k,i_1ji_4}
	+ \frac{Z_{k,ji_4}}{Z_{k,ij}}\partial_{\phi_{i_1}}Y_{k,ji_4}U^{(3)}_{k,i_2ii_3} \biggr] + \frac{1}{Z_{k,ij}} I_{3,k}^{i_1i_2i_3i_4}  U^{(3)}_{k,i_2ii_3}U^{(3)}_{k,i_1ji_4}  \biggr\},
\end{align}
with the notation $U^{(3)}_{i_1i_2i_3}=\partial_{\phi_{i_1}}U''_{k,i_2i_3}$, and we do not explicitly write the dependence on $\bd \phi$ and $\omega_n$. The parameter $\eta_{k,ij}= - \partial_t  \ln Z_{k,ij}$ is determined from the condition
\begin{equation}
	\int d^2\phi\, \partial_t Y_{k,ij}(i\omega_n=0,\bd \phi)=0 ,
\end{equation}
which is a direct consequence of~\eqref{eq:Y_constraint}. The threshold functions $I_{j,k}$ ($j=1,2,3$) are defined by
\begin{equation}
	\begin{split}
		&I_{1,k}^{i_1i_2}(\bd\phi)=\frac{1}{2\beta}\sum_{\omega_m}e^{-\omega_m^2/D^2}\tilde{\partial}_t G_{k,i_1i_2}(i\omega_m,\bd\phi),\\
		&I_{2,k}^{i_1i_2i_3i_4}(i\omega_n,\bd\phi)=-\frac{1}{\beta}\sum_{\omega_m}e^{-\omega_m^2/D^2}\tilde{\partial}_tG_{k,i_1i_2}(i\omega_m,\bd\phi)G_{k,i_3i_4}(i\omega_n+i\omega_m,\bd\phi),\\
		&I_{3,k}^{i_1i_2i_3i_4}(i\omega_n,\bd\phi)=\frac{I_{2,k}^{i_1i_2i_3i_4}(i\omega_n,\bd\phi)-I_{2,k}^{i_1i_2i_3i_4}(0,\bd\phi)}{\omega_n^2},
	\end{split}
\end{equation}
where the derivative operator $\tilde{\partial}_t=(\partial_t R_k)\partial_{R_k}$ acts only on the time dependence of $R_k$. Note that $I_{3,k}$ remains well defined in the limit $\omega_n \to 0$, which can be verified by expanding $I_{2,k}$ in powers of frequency. In practice, the flow equations are solved using the dimensionless quantities
\begin{equation}
	\omega_n=\frac{\omega_n}{k},\quad \tilde U_k(\bd\phi) =\frac{U_k(\bd\phi)}{k}, \quad \tilde Z_{k,ij}=kZ_{k,ij},\quad \tilde Y_{k,ij}(i\tilde\omega_n,\bd \phi)=Y_{k,ij}(i\omega_n,\bd \phi).
\end{equation}
Similarly to the potential [Eq.~\eqref{eq:Uharmonics}], the function $Y_{k,ij}(i\omega_n,\bd\phi)$ is expanded into a Fourier series.

\section[IV. Numerical results]{IV. Numerical results}

In this section, we present the numerical solution of the flow equations. We restrict ourselves to the symmetric case where $r_j=r$. Unless stated otherwise, all results are obtained within the BMW scheme with $N_0=1/2$. 

\subsection[A. Flows at $T=0$]{A. Flows at $T=0$}
\label{subsec:flowT0} 

For convenience, we decompose the effective potential into two parts: $U_k(\bd\phi)=V_k(\bd\phi)+U_k(0)$. $U_k(0)$ corresponds to the minimum value of the effective potential and determines the free energy $F=U_{k=0}(0)$. Since its flow does not influence other functions, it can be treated independently. Conversely, the flow of $V_k(\bd\phi)$ is coupled to that of the two-point vertex, i.e. $Y_k(i\omega_n,\bd\phi)$. 

At zero temperature, the Matsubara frequency $\tilde\omega_n\equiv \tilde\omega$ becomes a continuous variable. We find that the flow reaches a fixed point defined by the $k$-independent functions $\tilde V^*(\bd\phi)$, $\tilde Y^*(i\tilde\omega,\bd\phi)$ and the the $k$-independent parameter $\eta_{ij}=\eta$. For $K>3/2$, the fixed-point potential $\tilde V^*(\bd\phi)$ vanishes, in agreement with the scaling dimension $1-2K/3$ of the reflection coefficient $r$. The parameter $\eta$ takes the value $3-4K/3$ for $3/2\leq K \leq 9/4$ and vanishes for $K>9/4$. The function $\tilde Y^*(i\tilde\omega,\bd\phi)$ is nonzero but the self-energy $\tilde\Delta_k(i\tilde\omega,\bd\phi)\sim k^{1-\eta}\tilde\omega^2$ vanishes for $k\to 0$ since $\eta<1$. Therefore, the frequency dependence of the two-point vertex $\Gamma_{k=0}^{(2)}(i\omega,\bd\phi)$ is given by $|\omega|/\pi K$ in the small-frequency limit. 

Conversely, the flow reaches a nontrivial fixed point for $K<3/2$, defined by a nonzero effective potential $\tilde V^*(\bd\phi)\neq 0$, as shown in Fig.~\ref{fig:potential_r_nu} for the case $K=1$. For this value of the Luttinger parameter, the approach to the fixed point is characterized by the (correction-to-scaling) exponent $\nu\simeq 0.408$. This value is in good agreement with the exact value $\nu=2/5$ at the 3CK fixed point obtained from conformal field theory (CFT)~\cite{sm_affleck2001,sm_yi1998,sm_yi2002,sm_bao2017quantum}. The dimensionful potential $V_k=k\tilde V_k$ vanishes as $k$ approaches 0, which is a consequence of convexity and periodicity: The effective potential, $U_{k=0}$, is defined by a Legendre transform and is therefore a convex function. Since it is periodic, it must be constant, i.e., $V_{k=0}(\bd\phi)=U_{k=0}(\bd\phi)-U_{k=0}(0)=0$.

\begin{figure}
	\centering
	\includegraphics[width=18cm]{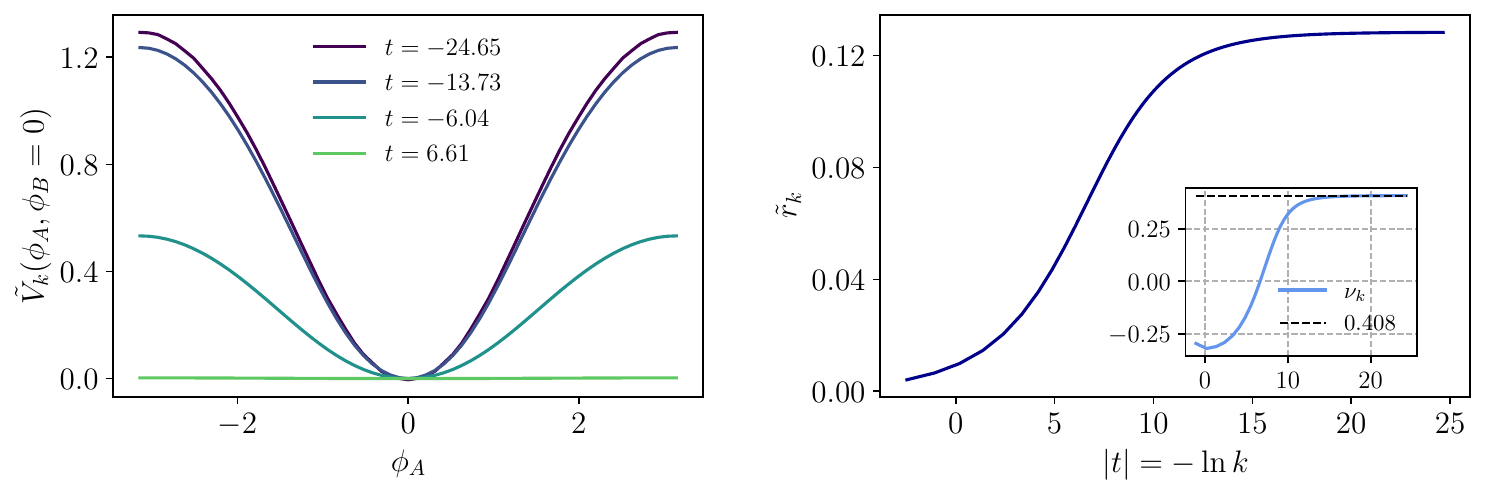}
	\vspace{-0.8cm}
	\caption{\label{fig:potential_r_nu} Flow of the dimensionless effective potential $\tilde V_k(\bd\phi)$ obtained for $K=1$. (Left) $\tilde V_k(\phi_A,\phi_B=0)$ as a function of $\phi_B$ for various RG times $t=\ln k$. For convenience, we use the variables $\phi_A=2\phi_2/\sqrt{3}$ and $\phi_B=\phi_1-\phi_2/\sqrt{3}$. (Right) Effective reflection coefficient $\tilde r_k=|\tilde r_k^{(1,0)}|=|\tilde r_k^{(0,1)}|=|\tilde r_k^{(1,1)}|$ as a function of $t$; see Eq.~\eqref{eq:Uharmonics}. The inset shows the running exponent $\nu_k=\partial_t \ln|\tilde r_k-\tilde r_{k=0}|$, which converges to $0.408$.}
\end{figure}	

When $K<3/2$, $\eta_k$ tends to the value $\eta=1$ for $k\to 0$: $Z_{k,ij}$ diverges as $1/k$ and $\tilde Z_{k,ij}$ reaches a fixed point value $\tilde Z^*_{ij}$. The self-energy $\Delta_k(i\omega,\bd\phi)$ behaves as $Z_{k,ij}\omega^2\sim k^{-\eta}\omega^2$ for $|\omega|\ll k$. For $|\omega|\gg k$, we instead find $\Delta_k(i\omega,\bd\phi)\simeq \Delta_{k=0}(i\omega,\bd\phi)\propto |\omega|^{2-\eta}\sim |\omega|$, as shown in Fig.~\ref{fig:Z2_Y} for $K=1$~\footnote{In the BMW approach, the same parameter $\eta$ controls the behavior of the self-energy in the range $|\omega|\ll k$, $\Delta_k(i\omega)\sim k^{-\eta}\omega^2$, and in the range $|\omega|\gg k$, $\Delta_k(i\omega)\sim |\omega|^{2-\eta}$~\cite{sm_benitez2012}.}. This is a remarkable result: Although the Luttinger parameter $K$ is not renormalized directly by the flow, the $|\omega|/\pi K$ term in the two-point vertex is corrected by the self-energy term $\Delta_{k=0}(i\omega,\bd\phi)$. As we will demonstrate in the next section, the linear frequency dependence of the self-energy $\Delta_{k=0}(i\omega,\bd\phi)$ is crucial for the emergence of a conductance value $G^*=G(\omega\to 0)$, which is intermediate between 0 and $e^2K/h$.

\begin{figure}[b]
	\centering
	\includegraphics[width=18cm]{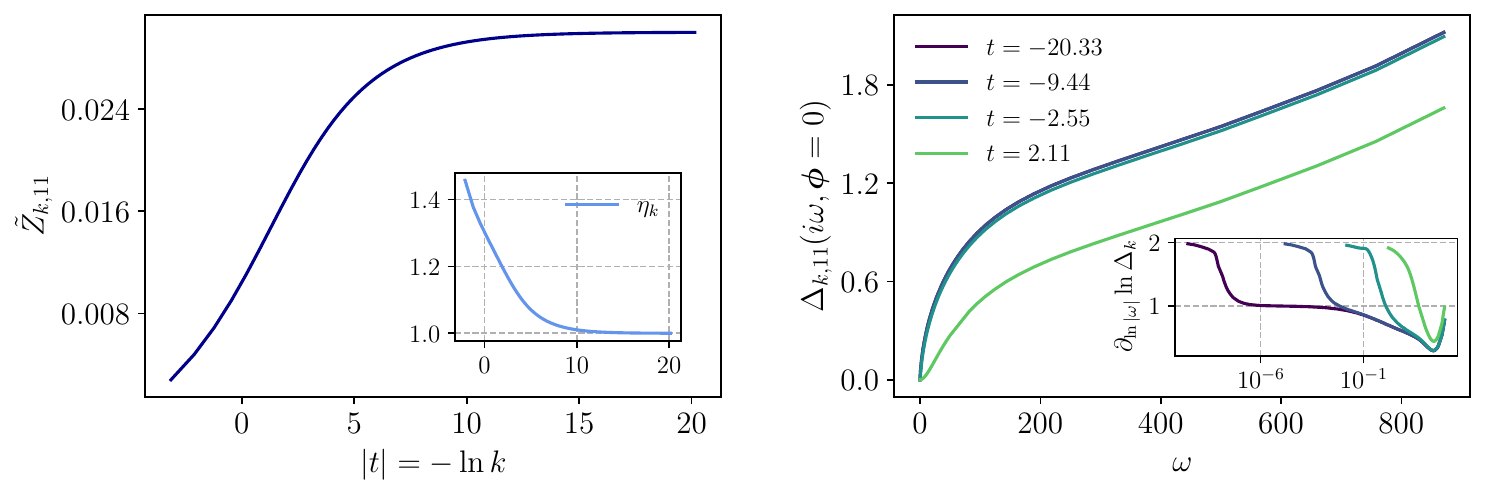}
	\vspace{-0.7cm}
	\caption{\label{fig:Z2_Y}   
		(Left) Flows of $\tilde Z_{k,11}$ and $\eta_{k,ij}=\eta_k$ (inset) for $K=1$. (Right) Self-energy $\Delta_{k,11}(i\omega,\bd\phi=0)$ obtained for various values of the RG time $t=\ln k$. The inset shows $\partial_{\ln|\omega|} \Delta_k(i\omega,\bd\phi=0)$ and the linear behavior $\Delta_k(i\omega,\bd\phi=0)\sim |\omega|$ emerging in the regime $|\omega|\gtrsim k$.}
\end{figure}

Capturing the full renormalization of the functions $\tilde{V}_k$ and $\tilde Y_k$ requires an increasingly large number of harmonics in the functions $\tilde V_k$ and $\tilde Y_k$ as $K$ decreases. This limits practical computations to $K\gtrsim 0.9$. 

\subsection[B. Universal conductance at $T=0$]{B. Universal conductance at $T=0$}
\label{subsec:universalG} 

By inverting the two-point vertex~\eqref{eq:Gamma_BMW}, we obtain the propagator $G_{k=0}(i\omega,\bd\phi)$ and, using~\eqref{eq:cond_symm}, the conductance 
\begin{align} \label{eq:cond1}
	G(\omega) &= \frac{e^2}{4\pi^2} {\rm Re } \left\{|\omega|\,
	{\rm tr} [G_{k=0}(i\omega,\bd\phi=0)]
	\bigr|_{i\omega\to \omega+i0^+} \right\} \nonumber\\ 
	&= \frac{e^2}{4\pi^2} {\rm Re } \left\{|\omega|\frac{\frac{2}{\pi K} |\omega| + \Delta_{k=0,11}(i\omega,0) + \Delta_{k=0,22}(i\omega,0)}{\bigl[\frac{|\omega|}{\pi K} + \Delta_{k=0,11}(i\omega,0)\bigr]\bigl[\frac{|\omega|}{\pi K} + \Delta_{k=0,22}(i\omega,0)\bigr]  - \Delta_{k=0,12}(i\omega,0) \Delta_{k=0,21}(i\omega,0)} \biggr|_{i\omega\to \omega+i0^+} \right\} .
\end{align}
We have used $U''_{k=0,ij}(0)=0$ since $U_{k=0}(\bd\phi)$ is independent of $\bd\phi$. The conductance $G^*=G(\omega=0)$ depends on the small-frequency behavior of the self-energy,  $\Delta_{k=0}(i\omega)\sim|\omega|^{2-\eta}$. For $K>3/2$, we have $2-\eta>1$, and we recover the expected result in the perfect transmission limit: $G^*=e^2K/2\pi\equiv e^2K/h$ (restoring Planck's constant). For $K<3/2$, $\eta=1$ and we obtain a nontrivial conductance $G^*$ with $0<G^*<e^2K/h$. For $K=1$, we find $G^*\simeq 0.677$ (in units of $e^2/h$), which is in good agreement with the exact result at the 3CK fixed point obtained from CFT, $G^*_{\rm 3CK}=2\sin^2(\pi/5)\simeq 0.691$~\cite{sm_affleck2001,sm_yi1998,sm_yi2002,sm_bao2017quantum}. Note that the limits $k \to 0$ and $\omega \to 0$ do not commute; the limit $k\to 0$ must always be taken first. This is the reason why the LPA, or a derivative expansion of the effective action, do not allow us to compute the conductance, as they provide the two-point vertex only in the limit $|\omega|\ll k$.  

\begin{figure}[b]
	\centering
	\includegraphics[width=9cm]{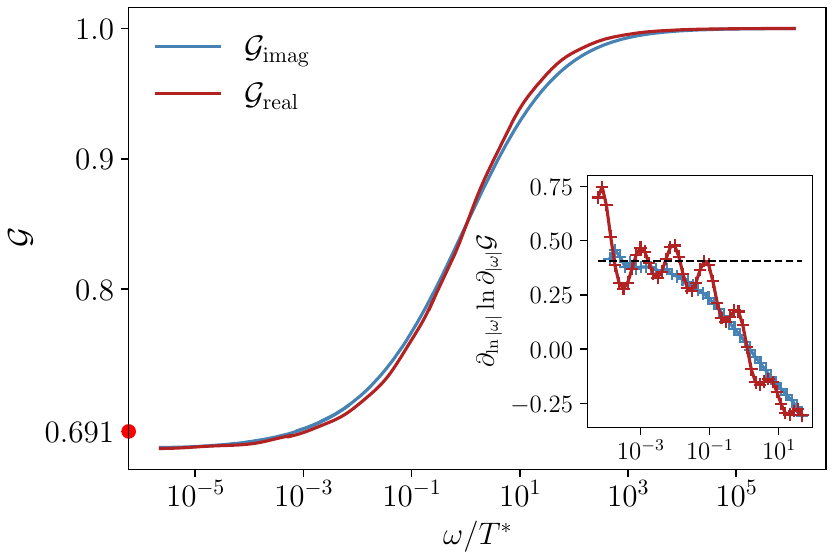}
	\vspace{-0.3cm}
	\caption{Universal scaling functions $\mathcal G_{\rm imag}$ and $\mathcal G_{\rm real}$ for $K=1$ [Eqs.~(\ref{Gimag},\ref{Greal})]. The exact result $\mathcal{G}(0)=2\sin^2(\pi/5)\simeq 0.691$ obtained from CFT is indicated by a red point on the vertical axis. The inset shows that the low-frequency behavior
		is compatible with $\mathcal G=\mathcal G^*+c|\omega|^\nu$, with the exponent $\nu\simeq 0.408$ (shown by the horizontal dashed line) obtained from the approach to the fixed point.}
	\label{fig:cond_real_imag}
\end{figure}

To obtain the frequency dependence of the conductance, we perform the analytic continuation in~\eqref{eq:cond1} using Pad\'e approximants~\cite{sm_rose2015}. The conductance can be written in the scaling form
\begin{equation} \label{Gimag} 
	G(i\omega) = \frac{e^2K}{h} {\cal G}_{\rm imag}\left(\frac{\omega}{T^*} \right) 
\end{equation}
as a function of the imaginary frequency $\omega$, or 
\begin{equation} \label{Greal} 
	G(\omega) = \frac{e^2K}{h} {\cal G}_{\rm real}\left(\frac{\omega}{T^*} \right) 
\end{equation}
after analytic continuation. Here, $T^*$ is a low-energy scale that characterizes the crossover between high- and low-frequency behaviors. It is defined such that $G(\omega=T^*)$  is halfway between the fully transparent value of $e^2K/h$ and the universal zero-frequency conductance $G^*(\omega=0)=\mathcal{G}(0)$. Its dependence on microscopic parameters can be determined from a perturbative analysis near the ballistic regime $r=0$ (see Appendix~\ref{sec_perturbation_rzero}). The universal scaling functions $\mathcal G_{\rm imag}$ and $\mathcal{G}_{\rm real}$ are nearly identical, as shown in Fig.~\ref{fig:cond_real_imag} for $K=1$. In the low-frequency limit, we expect $\mathcal G=\mathcal G^*+c|\omega|^\nu$, with $\nu$ the exponent of the leading irrelevant operator at the fixed point. Accurately determining $\nu$ is difficult due to a lack of numerical precision in the small-frequency limit. However, the result is compatible with the value $\nu\simeq 0.408$ obtained from the approach to the fixed point in Sec.~\ref{subsec:flowT0}; see the inset in Fig.~\ref{fig:cond_real_imag}. 

While the Pad\'e approximant method is known to be reliable for performing analytic continuation from numerical FRG data at $T=0$~\cite{sm_rose2015,sm_rose2016}, it usually fails at nonzero temperature. This is the main obstacle to obtaining the temperature-dependent conductance $G(\omega,T)$.

\subsection[C. Universal entropy]{C. Universal entropy}

\begin{figure}
	\centering
	\includegraphics[width=9.0cm]{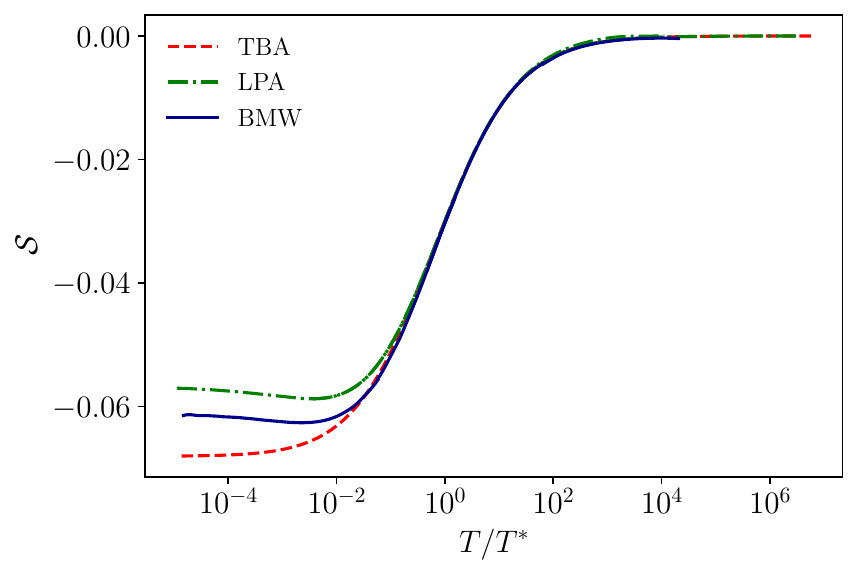}
	\vspace{-0.4cm}
	\caption{\label{fig:entropy_comparison} Universal entropy computed in the LPA and the BMW approximations, and compared with the exact TBA result.
	}
\end{figure}

The entropy of the metallic island is usually defined by~\cite{sm_han2022} 
\begin{equation}
	S_{\rm imp}(T) = S(T)- S(T)\bigl|_{N_0=0} , 
\end{equation}
i.e. by subtracting the entropy at $N_0=0$ from the entropy $S(T)=-\partial F/\partial T=-\partial U_{k=0}(0)/\partial T$ evaluated at $N_0=1/2$. For $K=1$, the exact value of the zero-temperature impurity entropy is known from CFT, $S_{\rm imp}(0)=\ln[(1+\sqrt{5})/2]$, and is interpreted in terms of Fibonacci anyons~\cite{sm_han2022,sm_lopes2020anyons,sm_komijani2020isolating,sm_lotem2022manipulating}. However, computing the entropy with FRG when $N_0$ differs from a half-integer number is difficult since the system flows to the strong-coupling (zero-transmission) fixed point and the $\tilde r_k^{(m,n)}$'s diverge when $k\to 0$. For this reason, we prefer to compute 
\begin{equation}
	\Delta S(T) = S(T)- S(T)\bigl|_{r=0} , 
\end{equation}
taking as reference point the purely ballistic fixed point. The exact CFT result is then given by \mbox{$\Delta S(0)=\ln[(1+\sqrt{5})/2\sqrt{3}]$} for $K=1$~\cite{sm_affleck2001}. To obtain the temperature dependence of the entropy from FRG, we compute the free energy $F=U_{k=0}(\bd\phi=0)$ and take a numerical derivative with respect to $T$. 

In the large-bandwidth limit, the entropy satisfies the scaling form 
\begin{equation}
	\Delta S(T) = \mathcal{S}\left( \frac{T}{T^*} \right) , 
\end{equation} 
where $\mathcal{S}$ is a universal scaling function and $T^*$ the characteristic energy scale introduced in Sec.~\ref{subsec:universalG}. Figure~\ref{fig:entropy_comparison} shows the function $\mathcal{S}$ obtained from the LPA and the BMW approximation for $K=1$, as well as the exact result obtained from the thermodynamic Bethe ansatz (TBA), whose calculation is detailed in Appendix~\ref{app:Bethe}. There is a very good agreement with TBA down to $T/T^*\simeq 5\times 10^{-1}$ for the LPA and $T/T^*\simeq 5\times 10^{-2}$ for the BMW approximation. However, at lower temperatures, the FRG becomes unreliable, and the slight increase in $\mathcal{S}$ as $T$ decreases is clearly an artifact. If we accept the hypothesis that the zero-temperature entropy $\Delta S(0)$ originates from Fibonacci anyons, then the failure of the FRG at low temperatures could indicate that this technique does not fully capture these excitations in the present context.

\appendix

\section[A. Numerical integration of the flow equations]{Appendix A: Numerical integration of the flow equations} 

In this appendix, we discuss the numerical integration of the flow equations, focusing on the BMW approximation. 

\subsection[1. Zero temperature]{1. Zero temperature}

When $T=0$, the Matsubara frequency $\omega_n\equiv \omega$ becomes a continuous variable, and the Matsubara sums are replaced by integrals. We follow the flow of the frequency-dependent dimensionless functions, e.g. $\tilde Y_{k,ij}(i\tilde\omega,\bd\phi)$, with $\tilde\omega$ belonging to a fixed, regular grid with the range $0\leq |\tilde{\omega}|\leq \tilde\omega_{\rm max}$. With our choice of cutoff function, we find that the contribution of the frequencies beyond $\tilde{\omega}_{\rm max}=6$ is extremely small and is negligible. When computing $I_{2,k}$, one must evaluate $\tilde{Y}_k$ for frequencies larger than $\tilde\omega_{\rm max}$. In such cases, following~\cite{sm_benitez2012}, we set $\tilde{Y}_k(i\tilde\omega,\bd\phi)=\tilde Y_k(i\tilde\omega_{\rm max},\bd\phi)$. 

However, determining $\tilde Y_k(i\tilde\omega,\bd\phi)$ and $\tilde\Delta_k(i\tilde\omega,\bd\phi)$ is insufficient. One must also compute the dimensionful self-energy $\Delta_k(i\omega,\bd\phi)$ to obtain the conductance. For a given (positive) frequency $\omega$, the dimensionless frequency $\tilde\omega=\omega/k$ remains in the (fixed) $\tilde\omega$-grid as long as $k\geq\omega/\tilde\omega_{\rm max}$. We must then switch from dimensionless to dimensionful formulation of the flow equations. It turns out to be crucial to continue the flow in the dimensionful form to obtain accurate results for the 3CK fixed point. Since the frequency dependence of the self-energy is linear rather than quadratic for $|\omega|\gtrsim k$, it is convenient to introduce the function $X_{k,ij}$ such that $\Delta_{k,ij}(i\omega,\bd\phi)=|\omega|X_{k,ij}(i\omega,\bd\phi)$. Its BMW flow equation is
\begin{align}
	\partial_t X_{k,ij}= &\sum_{i_1,i_2}I_{1,k}^{i_1i_2} \partial_{\phi_{i_1}}\partial_{\phi_{i_2}}X_{k,ij}
	+\sum_{i_1,i_2,i_3,i_4} \biggl\{ \frac{1}{|\omega|} [I_{2,k}^{i_1i_2i_3i_4}(i\omega)-I_{2,k}^{i_1i_2i_3i_4}(0)] U^{(3)}_{k,i_2ii_3}U^{(3)}_{k,i_1ji_4}\nonumber\\ & + 
	I_{2,k}^{i_1i_2i_3i_4}(i\omega) \bigl[  |\omega|\partial_{\phi_{i_2}}X_{k,ii_3}\partial_{\phi_{i_1}}X_{k,ji_4} + \partial_{\phi_{i_2}}X_{k,ii_3}U^{(3)}_{k,i_1ji_4}+\partial_{\phi_{i_1}}X_{k,ji_4}U^{(3)}_{k,i_2ii_3} \bigr]
	\biggr\}   .  
\end{align}
In this expression, $I_{2,k}(i\omega,\bd\phi)$ must be evaluated outside the  dimensionless-frequency grid when $|\omega|>k\tilde\omega_{\rm max}$. In that case, it is approximated by $I_{2,k}^{i_1i_2i_3i_4}(i\omega,\bd\phi)\simeq-2G_{k,i_3i_4}(i\omega,\bd\phi) I_{1,k}^{i_1i_2}(\bd\phi)$ using the approximation $G(i\omega+i\omega',\bd\phi)\simeq G(i\omega,\bd\phi)$ for $|\omega'|\leq k\tilde\omega_{\rm max}\leq |\omega|$~\cite{sm_benitez2009}. The dimensionless and dimensionful parts of the flow are related by the condition 
\begin{equation}
	X_{k,ij}(i\omega,\bd\phi)=\tilde Z_{k,ij} \tilde\omega_{\rm max} [ 1+\tilde Y_{k,ij}(i\tilde\omega,\bd\phi) ]
\end{equation}
at $k=\omega/\tilde{\omega}_{\rm max}$.

We find that  a Fourier expansion adapted to the triangular lattice in $\bd\phi$ (as in Eq.~\eqref{eq:Uharmonics}), a five-point finite differentiation in frequency, combined with the Simpson rule to compute the integrals, and an adaptive-step Runge-Kutta method produce stable and converging results. The dimensionless-frequency grid is defined in the range $[0,\tilde\omega_{\rm max}=6]$ with a step $\delta\tilde\omega=0.1$. The $\bd\phi$ grid is made of $35\times 35$ harmonics for the results presented. Note that when truncating the number of harmonics, care must be taken to preserve the triangular symmetry of the functions.

The main difficulty in numerically integrating the flow equations is related to the number of harmonics in the field expansion of the functions $\tilde U_k(\bd\phi)$ and $\tilde Y_k(i\tilde\omega,\bd\phi)$. The harmonic $\tilde r_k^{(m,n)}$ of rank $(m,n)$ in the effective potential~\eqref{eq:Uharmonics} has a scaling dimension $1-(2K/3)(m^2+n^2-mn)$. High-order harmonics are therefore strongly irrelevant and can be ignored (a similar analysis can be made for $\tilde Y_k(i\tilde\omega,\bd\phi)$). For $K\gtrsim 0.9$, we find that $35\times 35$ harmonics suffice to ensure the convergence of the functions $\tilde U_k$ and $\tilde Y_k$ to their fixed-point values $\tilde U^*$ and $Y^*$ as $k\to 0$. However, as $K$ decreases below $0.9$, a larger number of harmonics are necessary to ensure convergence, thus making the numerical integration of the flow equations more and more time consuming. We believe that values of $K$ smaller than 0.9 could be reached by allocating more numerical resources. On the other hand, the regime $K<2/3$ is likely to remain inaccessible since we expect a fully strong-coupling (zero-transmission) fixed point with diverging values of the coefficients $\tilde r_k^{(m,n)}$ as $k\to 0$. 

We estimate the accuracy of the FRG results from their sensitivity to the parameters $\gamma_1$ and $\gamma_2$ entering the regulator function $R_k$. The optimal values of $\gamma_1$ and $\gamma_2$ are chosen by minimizing the sensitivity of $G^*=G(\omega=0)$ to the regulator (principle of minimal sensitivity)~\cite{sm_canet2003,sm_balog2019,sm_DePolsi2022pms}, i.e., from the condition $dG^*/d\gamma_i=0$. The universal conductance $G^*$ at the 3CK fixed point changes by no more than three percent when $\gamma_1$ is varied in the interval $[0.7, 6]$ and $\gamma_2$ in the interval $[2,15]$. This three percent variation is similar to the two percent difference between the FRG estimate $G^*\simeq 0.677$ and the exact CFT result $G^*_{\rm 3CK}=2\sin^2(\pi/5)\simeq 0.691$.

\subsection[2. Non-zero temperature]{2. Non-zero temperature}

When $T \neq 0$, the dimensionless Matsubara frequencies $\tilde{\omega}_n = 2\pi n T / k$ form a quasi-continuum at the beginning of the RG flow, since the initial cutoff $k_{\rm in}$ is chosen much larger than all physical energy scales (in particular, $k_{\rm in}\gg T$). Consequently, the flow can be approximated by the zero-temperature flow as long as $k_{\rm in}$ is larger than $k_T = 2\pi n_{\rm max} T / \tilde{\omega}_{\rm max}$, where $n_{\rm max}$ is typically in the range $50-100$. When $k<k_T$, the Matsubara sums in the threshold functions cannot be approximated by integrals and must be evaluated explicitly. At this point, we switch to the dimensionful version of the flow, in which the threshold functions are evaluated as discrete Matsubara sums.

\section[B. Perturbative expansion near the ballistic fixed point]{Appendix B: Perturbative expansion near the ballistic fixed point} 
\label{sec_perturbation_rzero} 

In this section, we discuss the perturbative analysis near the ballistic fixed point $r=0$ and the high-frequency behavior of the conductance.

\subsection[1. Mobility]{1. Mobility}

The conductance tensor
\begin{equation}
	G_{ij}(i\omega_n) = \frac{e^2K}{2h} \bd G^T_i \cdot \mu(i\omega_n) \cdot \bd G_j  	
\end{equation}
can be expressed in terms of the mobility, defined by 
\begin{equation}\label{eq:mob_1}
	\mu_{ij}(i\omega_n)=\frac{|\omega_n|}{\pi K}\langle\varphi_i(i\omega_n)\varphi_j(-i\omega_n)\rangle .
\end{equation}
For the QBM described by the action~\eqref{eq:action_QBM}, the mobility gives the (average) long-time velocity of the particle subjected to an external force.  

To obtain an expression which is more convenient for perturbation theory, we consider the QBM action,
\begin{align}
	S[\bd\varphi,\bd h] &= \sum_{\omega_n}\frac{|\omega_n|}{2\pi K} \bigl[ |\bd\varphi(i\omega_n)|^2 - \bd h(-i\omega_n)\cdot \bd\varphi(-i\omega_n) \bigr] - \int_0^\beta d\tau \sum_a r_a \cos\Bigl(\bd G_a\cdot \bd\varphi-\frac{\pi}{3}\Bigr)  \nonumber\\ 
	&= \sum_{\omega_n}\frac{|\omega_n|}{2\pi K} \left[ |\bd\varphi(i\omega_n)|^2 - \frac{1}{4} |\bd h(-i\omega_n)|^2 \right] - \int_0^\beta d\tau \sum_a r_a \cos\Bigl(\bd G_a\cdot \Bigl(\bd\varphi+ \frac{\bd h}{2} \Bigr)-\frac{\pi}{3}\Bigr) 
\end{align}
in the presence of an external source. We have set $N_0=1/2$ and shifted the field $\bd\varphi\to \bd\varphi+\bd h/2$ to obtain the second line. We then obtain
\begin{align}\label{eq:mob_2}
	\mu_{ij}(i\omega_n) ={}& \frac{4\pi K}{|\omega_n|}  \frac{\delta^2\ln \mathcal{Z}[\bd h]}{\delta h_i(i\omega_n)\delta h_j(-i\omega_n)}\biggl|_{\bd h=0} \nonumber\\ 
	={}& \delta_{ij}
	- \frac{\pi K}{|\omega_n|} \sum_a r_a G_a^i G_b^j \Bigl\langle \cos\Bigl(\bd G_a\cdot \bd\varphi(0)-\frac{\pi}{3}\Bigr) \Bigr\rangle \nonumber\\ 
	& + \frac{\pi K}{|\omega_n|} \sum_{a,b}r_a r_b G_a^i G_b^j\int_0^\beta d\tau\,e^{i\omega_n \tau} \Bigl\langle\sin\Bigl(\bd G_a\cdot \bd\varphi(\tau)-\frac{\pi}{3}\Bigr)\sin\Bigl(\bd G_b\cdot \bd\varphi(0)-\frac{\pi}{3}\Bigr)\Bigr\rangle .
\end{align}

\subsection[2. Perturbative expansion to order $r^2$]{2. Perturbative expansion to order $r^2$}

We compute the mobility in a perturbative expansion with respect to the reflection coefficients $r_a$. We use the fact that the expectation value 
\begin{equation}
	\langle e^{i \sum_i \bd G_i \cdot \bd\varphi(\tau_i)} \rangle_0
\end{equation} 
is nonzero only if $\sum_i\bd G_i=0$. Here, $\langle\cdots\rangle_0$ denotes the average with respect to the ballistic action ($r_a=0$). We easily obtain 
\begin{equation} 
	\Bigl\langle \cos\Bigl(\bd G_a\cdot \bd\varphi(0)-\frac{\pi}{3}\Bigr) \Bigr\rangle 
	= \frac{r_a}{2} \int_0^\beta d\tau\,
	\bigl\langle e^{i \bd G_a\cdot [\bd\varphi(0) - \bd\varphi(\tau)] } \bigr\rangle_0
\end{equation}
and 
\begin{equation}
	\Bigl\langle\sin\Bigl(\bd G_a\cdot \bd\varphi(\tau)-\frac{\pi}{3}\Bigr)\sin\Bigl(\bd G_b\cdot \bd\varphi(0)-\frac{\pi}{3}\Bigr)\Bigr\rangle = \frac{\delta_{ab}}{2} \bigl\langle e^{i \bd G_a\cdot [\bd\varphi(0) - \bd\varphi(\tau)] } \bigr\rangle_0
\end{equation}
to leading order, which leads to 
\begin{equation} 
	\mu_{ij}(i\omega_n)=\delta_{ij}-\frac{\pi K}{|\omega_n|}\sum_a\frac{r_a^2}{2} G_a^i G_a^j\int_0^\beta d\tau\,(1-e^{i\omega_n\tau})\langle e^{i\bd G_a\cdot [\bd\varphi(0)-\bd\varphi(\tau)]}\rangle_{0} 
\end{equation}
to quadratic order in $r_a$. In the symmetric case ($r_a=r$), this yields the QPC conductance 
\begin{equation}\label{eq:mob_pert}
	G(i\omega_n)= \frac{e^2K}{h} \biggl\{ 1 -\frac{\pi K r^2}{|\omega_n|}\int_0^\beta d\tau\,(1-e^{i\omega_n\tau})\langle e^{i\bd G_1\cdot [\bd\varphi(0)-\bd\varphi(\tau)]}\rangle_{0}  \biggr\} ,
\end{equation}
where we have restored Planck's constant. 

\subsection[3. High-frequency behavior and crossover temperature $T^*$]{3. High-frequency behavior and crossover temperature $T^*$}

In this section, we focus on the zero-temperature case. Using
\begin{equation}
	\langle\varphi_i(0)^2 \rangle_{0} - \langle\varphi_i(\tau)\varphi_i(0)\rangle_{0}
	=\pi K \int_{-\infty}^{\infty} \frac{d\omega}{2\pi}e^{-\omega^2/D^2} \frac{1-e^{i\omega\tau}}{|\omega|}
	\simeq K \ln|D\tau| \quad (D|\tau|\gg 1),
\end{equation}
we obtain
\begin{equation}
	\langle e^{i G_1\cdot (\bd\varphi(0)-\bd\varphi(\tau))}\rangle_{0} 
	= e^{-\frac{1}{2} \langle \{\bd G_1\cdot [\bd\varphi(\tau)-\bd\varphi(0)]\}^2 \rangle} 
	\sim |D\tau|^{-4K/3} 
\end{equation}  
in the long-time limit $|\tau|\gg 1/D$. We deduce
\begin{equation}
	\int d\tau\,(1-e^{i\omega\tau})\langle e^{i\bd G_1\cdot [\bd\varphi(0)-\bd\varphi(\tau)]}\rangle_{0} \sim \frac{|\omega|^{4K/3-1}}{D^{4K/3}} 
\end{equation}
and 
\begin{equation} \label{eq:T_star_app}
	\frac{e^2K}{h} - G(i\omega) \sim r^2 \biggl| \frac{\omega}{T^*} \biggr|^{4K/3-2} \quad \mbox{with} \quad  
	\frac{T^*}{D} \propto \left( \frac{r}{D} \right)^{\frac{1}{1-2K/3}} .
\end{equation}
Equation~(\ref{eq:T_star_app}) is valid in the high-frequency limit $|\omega|\gg T^*$ and for $K<3/2$, where $G(i\omega) \simeq e^2K/h$. The power-law dependence of $T^*/D$ on $r/D$ is in perfect agreement with the FRG calculation, as shown in Fig.~\ref{fig:T_star} for $K=1$.  

\begin{figure}
	\centering
	\includegraphics[width=7.5cm]{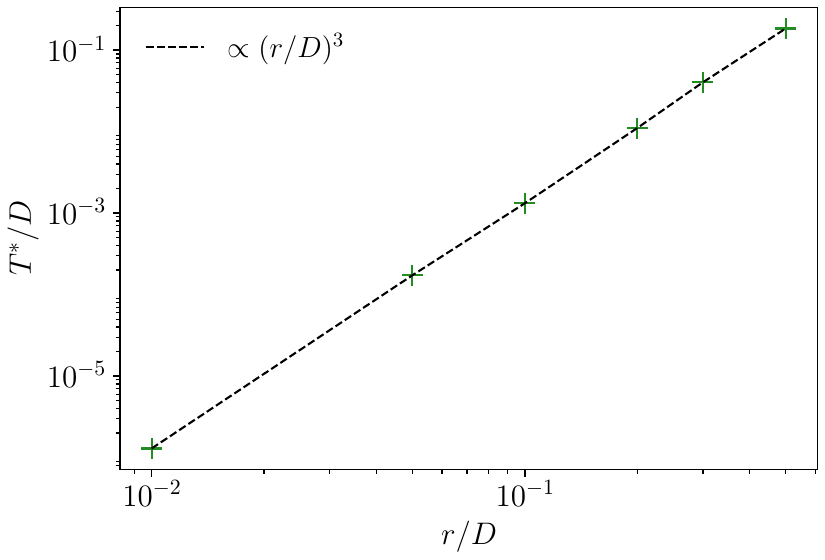}
	\vspace{-0.3cm} 
	\caption{\label{fig:T_star} Crossover scale $T^*$ as a function of the reflection coefficient $r$ for $K=1$.  The dashed line shows the power-law dependence $T^*/D \propto (r/D)^3$ [Eq.~\eqref{eq:T_star_app}].}
\end{figure}

\section[C. Thermodynamic Bethe ansatz for $K=1$]{Appendix C: Thermodynamic Bethe ansatz for $K=1$}
\label{app:Bethe}

The QBM model on a triangular lattice defined by the action~\eqref{eq:action_QBM} is integrable when $K=1$. Although a numerical solution of the TBA equations is not found in the literature, they are given in Sec.~VI.A of Ref.~\cite{sm_affleck2001},
\begin{equation} \label{eq:tba}
	\begin{split} 
		\epsilon_1(\theta)&=m_1e^\theta-T \int_{-\infty}^\infty \frac{d\theta'}{2\pi}\frac{1}{\cosh (\theta-\theta')}\ln(1+e^{-\epsilon_2(\theta')/T}) , \\
		\epsilon_2(\theta)&=m_2e^\theta-T\int_{-\infty}^\infty \frac{d\theta'}{2\pi}\frac{1}{\cosh (\theta-\theta')}\ln[(1+e^{-\epsilon_1(\theta')/T})(1+e^{-\epsilon_3(\theta')/T})] , \\
		\epsilon_3(\theta)&=m_3e^\theta-T\int_{-\infty}^\infty \frac{d\theta'}{2\pi}\frac{1}{\cosh (\theta-\theta')}\ln(1+e^{-\epsilon_2(\theta')/T})  . 
	\end{split} 
\end{equation}
These equations are expressed in terms of the pseudo-energies $\epsilon_i$. The mass parameters $m_i$ are complicated functions of the microscopic parameters $r_i$, but their exact expression is not necessary to determine the universal entropy crossover. The impurity free energy is given by
\begin{align}
	F_{\rm imp}(T)=-T\int_{-\infty}^\infty \frac{d\theta}{2\pi}\frac{1}{\cosh(\theta-\theta_B)}\ln(1+e^{-\epsilon_1(\theta)/T}) , 
\end{align}
where $T_B\propto e^{\theta_B}$ is also a complicated function of the coefficients $r_i$. Introducing $f_i(\theta)=\epsilon_i(\theta+\theta_B)/T$, we can rewrite Eqs.~\eqref{eq:tba} in the form
\begin{equation}\label{eq:bethe_B}
	\begin{split} 
		f_1(\theta)&=\frac{m_1e^{\theta_B}}{T}e^\theta-\int_{-\infty}^\infty \frac{d\theta'}{2\pi}\frac{1}{\cosh (\theta-\theta')}\ln(1+e^{-f_2(\theta')}) , \\
		f_2(\theta)&=\frac{m_2e^{\theta_B}}{T}e^\theta-\int_{-\infty}^\infty \frac{d\theta'}{2\pi}\frac{1}{\cosh (\theta-\theta')}\ln[(1+e^{-f_1(\theta')})(1+e^{-f_3(\theta')})] , \\
		f_3(\theta)&=\frac{m_3e^{\theta_B}}{T}e^\theta-\int_{-\infty}^\infty \frac{d\theta'}{2\pi}\frac{1}{\cosh (\theta-\theta')}\ln(1+e^{-f_2(\theta')}) , 
	\end{split} 
\end{equation}
with
\begin{equation}
	F_{\rm imp}(T)=-T\int_{-\infty}^\infty \frac{d\theta}{2\pi}\frac{1}{\cosh (\theta)}\ln(1+e^{-f_1(\theta)}) . 
\end{equation}
Reference~\cite{sm_affleck2001} demonstrates that the crossover between the ballistic fixed point ($r = 0$) and the intermediate 3CK fixed point occurs when $m_1 = m_2 = 0$. Consequently, the parameters $m_i$ and $\theta_B$ appear only in the combination $m_3 e^{\theta_B} = T^*_{\rm TBA}$. The functions $f_i$ depend solely on the ratio $T/T^*_{\rm TBA}$, and it follows directly that \mbox{$\Delta S_{\rm TBA} = -\partial F_{\rm imp} / \partial T$} is a universal function of $T/T^*_{\rm TBA}$.

\endgroup  

\end{document}